\documentclass[a4paper,fleqn,usenatbib]{mnras}

\usepackage[normalem]{ulem}

\usepackage[T1]{fontenc}
\usepackage{ae,aecompl}

\usepackage{graphicx}   
\usepackage{amsmath}    
\usepackage{amssymb}    
\usepackage{rotating}   

\title[Properties of satellites in phase-space]{Physical properties of SDSS satellite galaxies
in projected phase-space}

\author[A. Pasquali et al.]{
A. Pasquali,$^{1}$\thanks{E-mail: pasquali@uni-heidelberg.de}
R. Smith,$^{2}$
A. Gallazzi,$^{3}$
G. De Lucia,$^{4}$
S. Zibetti,$^{3}$
M. Hirschmann$^{5}$
\and and S.~K. Yi$^{6}$
\\
$^{1}$ Astronomisches Rechen-Institut, Zentrum f\"ur Astronomie der Universit\"at Heidelberg,
M\"onchhofstr. 12 - 14, 69120 Heidelberg, Germany\\
$^{2}$ Korea Astronomy and Space Science Institute, 776, Daedeokdae-ro, Yuseong-gu, Daejeon 34055, Republic of Korea\\ 
$^{3}$ INAF - Osservatorio  Astrofisico di Arcetri, Largo Enrico Fermi 5, 50125 Firenze, Italy\\
$^{4}$ INAF - Osservatorio Astronomico di Trieste, Via Tiepolo 11, 34131 Trieste, Italy\\
$^{5}$ Sorbonne Universit\'e, UPMC-CNRS, UMR7095, Institut d'Astrophysique de Paris, 75014, Paris, France\\ 
$^{6}$ Department of Astronomy and Institute of Earth-Atmosphere-Astronomy, Yonsei University, Seoul 03722, Korea
}

\date{Accepted 2018 December 25. Received 2018 December 15; in original form 2018 August 8}

\pubyear{2018}

\begin{document}
\label{firstpage}
\pagerange{\pageref{firstpage}--\pageref{lastpage}}
\maketitle

\begin{abstract}
We investigate how environment affects satellite galaxies using their location within
the projected phase-space of their host haloes from the Wang et al.'s group catalogue.
Using the Yonsei Zoom in Cluster Simulations, we derive zones of constant mean infall
time $\overline{\rm T}$$_{\rm inf}$ in projected phase-space, and catalogue in which zone
each observed galaxy falls. Within each zone we compute the mean observed galaxy properties
including specific star formation rate, luminosity-weighted age, stellar metallicity and
[$\alpha$/Fe] abundance ratio. By comparing galaxies in different zones, we inspect
how shifting the mean infall time from recent infallers ($\overline{\rm T}$$_{\rm inf}~<$~3~Gyr)
to ancient infallers ($\overline{\rm T}$$_{\rm{inf}}>$~5~Gyr) impacts galaxy properties
at fixed stellar and halo mass. Ancient infallers  are more quenched, and the impact of environmental
quenching is visible down to low host masses ($\leq$~group masses). Meanwhile,
the quenching of recent infallers is weakly dependent on host mass, indicating they have
yet to respond strongly to their current environment. [$\alpha$/Fe] and especially
metallicity are less dependent on host mass, but show a dependence on
$\overline{\rm T}$$_{\rm{inf}}$. We discuss these results in the context of longer exposure
times for ancient infallers to environmental effects, which grow more efficient in hosts with a 
deeper potential well and a denser intracluster medium. We also compare our satellites with a 
control field sample, and find that even the most recent infallers ($\overline{\rm T}$$_{\rm{inf}}$~<~2 Gyr) 
are more quenched than field galaxies, in particular for cluster mass hosts. This supports
the role of pre-processing and/or faster quenching in satellites.
\end{abstract}

\begin{keywords}
galaxies: clusters: general -- galaxies: evolution -- galaxies: general -- galaxies: haloes -- galaxies: stellar content 
\end{keywords}


\section{Introduction}

It is now well accepted that environmental effects are acting to transform galaxies in their morphology (Dressler 1980), 
colours (e.g. Tanaka et al. 2004, Weinmann et al. 2006, Weinmann et al. 2009, Vulcani et al. 2015), atomic gas content 
(Giovavelli \& Haynes 1985, Chung et al. 2009, Jaff\'e et al. 2015, Jaff\'e et al. 2018), and their star formation rates 
(e.g. Hashimoto et al. 1998, Domin\'inguez et al. 2002, G\'omez et al. 2003, Kauffmann et al. 2004, Poggianti
et al. 2008, Balogh et al. 2011, McGee et al. 2011, Hou et al. 2013, Lin et al. 2014, Mok et al. 2014, Davies et al. 2016,
Lofthouse et al. 2017). A variety of mechanisms 
whose efficiency is environmentally dependent may be responsible for these transformations 
including: ram pressure stripping (Gunn \& Gott 1972), starvation (Larson et al. 1980, Balogh et al. 2000, Bekki 2009), 
harassment (Moore et al. 1996, Moore et al. 1998, Gnedin et al. 2004, Smith et al. 2010, 2012, 2015),
and galaxy-galaxy mergers (Barnes \& Hernquist 1996). Until fairly recently, many 
studies of galaxy environment have often focused on the densest environments such as clusters.  However, it is now clear that 
the transformations can begin to occur in lower density environments, and that galaxy colours and star formation rates 
systematically vary with the density of their environment over a density range from voids, to filaments and walls, through 
groups, all the way up to cluster densities (Cybulski et al. 2014). Furthermore, thanks to the hierarchical manner in which 
structure growth occurs, many galaxies which are members of clusters today may have previously spent time in a group 
environment, and suffered group pre-processing (Zabludoff \& Mulchaey 1998, Balogh et al. 2000, De Lucia et al. 2012, 
Hoyle et al. 2012, Vijayaraghavan \& Ricker 2013, Hirschmann et al. 2014), meaning they have experienced both group 
and cluster environmental effects. 
Recent studies have in fact revealed that galaxies in groups are more deficient in atomic gas 
(Brown et al. 2017). By stacking X-ray images of individual galaxies, Anderson et al. (2013, 2015) have revealed that 
on average all galaxies with masses down to $\sim$10$^{10.5}$ M$_{\odot}$ are surrounded by an extended X-ray emitting 
ionised gas that could give rise to starvation or ram pressure.
   
When trying to understand the impact of environment on galaxies, it is crucial to first control for the effect
of galaxy mass. 
This is because it is generally accepted that `mass quenching' is a crucial process controlling galaxy evolution. It can be 
broadly understood in terms of the `halo quenching' process (Gabor \& Dav\'e 2015), where the gas in haloes more massive than 
10$^{12}$ M$_{\odot}$ is hindered from cooling as it becomes shock-heated (Birnboim \& Dekel 2003; Dekel \& Birnboim 2006). 
However, galaxy simulations tend to agree that AGN feedback is an additional yet necessary process to keep galaxies quenched 
over long time-scales (e.g. Granato et al. 2004, Bower et al. 2006). Galaxy mergers (Hopkins et al. 2006) 
and/or disk instabilities (Dekel et al. 2009) have also been recognized as further quenching mechanisms. The effects of all these 
processes scale with galaxy mass, and thus must be disentangled from the effects of environment. This was clearly 
demonstrated in Peng et al. (2010), who showed that galaxy colour is a function of both galaxy mass and local environmental 
density, and by Pasquali et al. (2010), who found that satellite stellar age and metallicity depend on both galaxy mass and 
host halo mass.  

When a galaxy first enters an environment of different density (e.g. a field galaxy falls into the cluster), clearly we should 
expect it to take a finite time for its new environment to take effect. Recently, the time it takes for galaxies to be fully 
quenched has been a subject of much discussion in the literature, with the general conclusion that the transformation takes 
place over several gigayears (cf. De Lucia et al. 2012, Wetzel et al. 2013, Oman \& Hudson 2016, Hirschmann et al. 2014). 
Such long time scales 
for environmentally-induced quenching are supported also by the observations that the relation between the element abundance 
ratio [$\alpha$/Fe] and stellar mass is to a large extent independent of galaxy hierarchy and halo mass, as we find in Gallazzi 
et al. (2018, in prep). Simply measuring the local environmental density may 
not provide all the relevant information, as galaxy properties would be expected to vary as a function of time spent in that 
environment. In other words, at fixed environmental density, there will be some galaxies that are very recent newcomers to that 
environment and whose properties may be much more shaped by their previous environment. This issue may be more severe in 
regions where there has been rapid recent growth, causing the environment to be flooded by recent infallers. Ideally, we 
would greatly benefit from being able to control for the effect of infall time - the time at which an object joined its 
environment.
   
One promising tool that may allow us to have some handle on the infall times of satellites into their current host is 
through the use of phase-space diagrams, combining cluster-centric (or group-centric) velocity with cluster-centric 
(or group-centric) radius. As a function of distance alone, the infall time is poorly constrained. This is because, in 
$\Lambda$CDM, typically the orbits of an infalling substructure are very plunging and eccentric (Gill et al. 2005, 
Wetzel 2011). 
As a result, at a fixed radius inside a cluster there may be objects falling in for the first time, combined with objects 
that have been orbiting within the cluster for many gigayears, creating a wide distribution of infall times. 
As a consequence, any correlation that can be recognized between mean infall time and cluster-centric distance suffers
from a large scatter (see Gao et al. 2004, De Lucia et al. 2012).
However, when 
cluster-centric distance is combined with cluster-centric velocity (e.g. a phase-space diagram), recent infallers to the 
cluster tend to shift to higher velocities than objects that have been in the cluster for some time. Thus different regions 
of phase-space tend to be dominated by objects with a narrower range of infall time. In a 3D phase-space diagram (using the 
3D cluster-centric radius and 3D cluster-centric velocity), different infall time populations separate quite neatly  
(e.g. the left panel of Figure 2 of Rhee et al. 2017). Nevertheless, when we consider observable quantities 
(line of sight velocity and projected radius from the cluster), there is often much more mixing of objects with differing 
infall times as a result of projection effects. Despite this, the shape of the infall time distribution systematically varies
with location in projected phase-space (Oman \& Hudson 2016, see for example Figure 5 of Rhee 2017).
   
A number of authors have used phase-space diagrams as tools to better understand galaxy evolution, including 
Mahajan et al. (2011, star formation), Muzzin et al. (2014, high redshift quenching), Hern\'andez-Fern\'andez et al. (2014) and
Jaff\'e et al. (2015, gas fractions and ram pressure), Oman \& Hudson (2016, quenching timescales), Yoon et al. (2017, ram 
pressure in Virgo), Jaff\'e et al. (2018, jelly fish galaxies undergoing ram pressure). In this paper, in order to take the next 
step forward, we will attempt to control for the effect of galaxy mass 
and also galaxy environment, while simultaneously controlling for the effect of infall time of the galaxies into their 
environment 
using projected phase-space diagrams. Since it is impossible to assign each galaxy an infall time, due to the 
mixing of infall times in each region of a projected phase-space diagram, we will theoretically define various zones in 
phase-space such that we can assume, by comparison with cosmological simulations of clusters and groups, that the mean of 
the infall time distribution is changing systematically with zone. Then, by comparing the population of galaxies 
in each zone, we can study how galaxy mean, observed properties change as their mean infall time shifts with zone. 
Our sample includes satellite galaxies with masses varying from dwarfs to giants, and host haloes with masses varying 
from large galaxies up to massive 
clusters. By spectral fitting, we can study multiple properties of our galaxies including luminosity-weighted age, specific 
star formation rate, metallicity and $\alpha$-abundances. Indeed this paper represents the first attempt to investigate 
how all of these properties vary as a function of galaxy mass, host mass while deliberately shifting the mean infall time of the 
population in a systematic way.

In Section 2 we describe the group catalogue from which our host and satellite samples are drawn, describe how spectral 
fitting is conducted, and summarise the working sample. In Section 3 we describe the cosmological simulations that we compare 
with, and explain how we choose the different zones in phase-space. In Section 4 we present our results for how galaxy 
properties vary with zone and compare our sample of satellite galaxies with a field sample. We discuss and interpret our 
results in Section 5, compute quenching time-scales as a function of stellar and host mass, and comment on 
possible indications of pre-processing. Finally, in Section 6 we summarise our results and draw conclusions.

\section{Data}
\subsection{Assessing galaxy environment}
We used the galaxy group catalogue drawn from the Sloan Digital Sky Survey Data Release 7 (SDSS DR7) by Wang et al. (2014) 
following the procedure of Yang et al. (2007). Briefly, the adaptive halo-based group finder developed
by Yang et al. (2005) was applied to all galaxies in the Main Galaxy Sample of the New York University
Value-Added Galaxy Catalogue (Blanton et al. 2005) for DR7 (Abazajian et al. 2009), which have an extinction 
corrected apparent magnitude brighter than $r = $ 17.72 mag, and are in the redshift range 0.01 $\leq z \leq$ 0.20 
with a redshift completeness C$_z >$ 0.7. This algorithm employs the traditional firedns-of-firends method with small linking
lengths in redshift space to sort galaxies into tentative groups and estimate the groups' characteristic stellar
mass and luminosity. In the first iteration, the adaptive halo-based group finder applies
a constant mass-to-light ratio of 500 $h$M$_{\odot}$/L$_{\odot}$ to estimate a tentative halo mass for each group. 
This mass is then used to evaluate the size and velocity dispersion of the halo embedding the group, which in turn 
are utilized to define group membership in redshift space. At this point a new iteration begins, whereby the group
characteristic luminosity and stellar mass are converted into halo mass using the halo occupation model of 
Yang et al. (2005). This procedure is repeated until no more changes occur in the group membership. 
\par\noindent
When applied to SDSS DR7, the adaptive halo-based group finder delivers three group samples: sample I, which relies only 
on those galaxies (593,736) with a spectroscopic redshift measured by SDSS; sample II, which extends the previous one by 3,115 
galaxies with SDSS photometry whose redshifts are known from alternative surveys; sample III, containing an additional 
36,602 galaxies without redshift because of fiber collision, but which were assigned the redshift of their closest 
neighbour (cf. Zehavi et al. 2002). 
In each group sample, galaxies are distinguished between centrals (the most massive
group members in terms of stellar mass M$_{\star}$), and satellites (all other group members less massive
than their group central). Stellar masses have been calculated using the relations between stellar mass-to-light
ratio and colour from Bell et al. (2003), while the dark matter masses M$_{\rm h}$ associated with the host groups 
have been estimated on the basis of the ranking of both the group total characteristic luminosity and the group
total characteristic stellar mass (cf. Yang et al. 2007 for more details). The two available M$_{\rm h}$ values
are reasonably consistent with each other (with an average scatter decreasing from 0.1 dex at the low-mass end to 0.05 dex
at the massive end), although More et al. (2011) showed that stellar mass constrains M$_{\rm h}$ better than luminosity. 
The method of Yang et al. (2007) is able to assign a value of M$_{\rm h}$ only to galaxy
groups more massive than $\sim$10$^{12}$ M$_{\sun}h^{-1}$ and containing at least one member with $^{0.1}$M$_r$ - 5log$h 
\leq$ -19.5 mag, where $^{0.1}$M$_r$ is the galaxy absolute magnitude in $r$ band corrected to $z =$ 0.1. For smaller
haloes, M$_{\rm h}$ has been extrapolated from the relations between the luminosity/stellar mass of central
galaxies and the halo mass of their host groups, thus reaching a limiting M$_{\rm h}$ of $\sim$10$^{11}$ 
M$_{\sun}h^{-1}$ (cf. Yang, Mo \& van den Bosch 2008). As shown by Yang et al. (2007), the typical uncertainty on M$_{\rm h}$
varies between $\sim$ 0.35 dex in the range M$_{\rm h}$ = 10$^{13.5}$ - 10$^{14}$ M$_{\odot}h^{-1}$ and $\sim$ 0.2 dex at
lower and higher halo masses. 

For our analysis of the observed galaxy properties in projected phase-space we made use of sample II, and  
halo masses derived from the group characteristic stellar mass. From sample II we extracted all galaxy groups with at 
least 4 members (the central galaxy plus three satellites). We choose this minimum number of members in order to ensure 
that our groups can truly be considered groups and not simply pairs of interacting galaxies.
Groups with 4 members may have a less well determined centre and thus the projected cluster-centric distances of their
satellites may be more uncertain. This mostly concerns the lower M$_{\rm h}$ haloes (10$^{12}$ - 10$^{13}$ M$_{\sun}h^{-1}$)
where the fraction of satellites living in groups with 4 members is 44$\%$ for 9 $\leq$ log(M$_{\star}$ M$_{\sun}h^{-2}) <$ 10,
59$\%$ for 10 $\leq$ log(M$_{\star}$ M$_{\sun}h^{-2}) <$ 10.5 and 72$\%$ at higher stellar masses. Nevertheless, we believe that
increasing the galaxy statistics by combining all groups in the low M$_{\rm h}$ range partially mitigates the effect of an uncertain
group centre on our results, and we will show that trends with phase-space zone still persist.  
\par\noindent
We used the spectroscopic redshifts of the satellites belonging to the same group to derive their line-of-sight velocities,
and to compute the velocity dispersion along the line of sight ($\sigma$) of their 
group using the standard unbiased estimator for the variance. 
We also determined the peculiar line-of-sight velocity of each satellite, $| \Delta {\rm V} |$, as the absolute value of the 
difference between its line-of-sight velocity and the mean of the satellite line-of-sight velocities.
The projected distance of each satellite from the luminosity weighted centre of
its host group is given in sample II as R$_{\rm proj}$ in kpc~$h^{-1}$. In addition, we calculated the group virial radius 
R$_{\rm 200}$ (at which the average group density is 200 times higher than the critical density) using the relation
(cf. Yang et al. 2007):

\begin{equation}
{\rm R_{\rm 200} [kpc~h^{-1}] = \frac{258.1 \times (M_{\rm h}/10^{12})^{1/3} \times (\Omega_{\rm m}/0.25)^{1/3}}{(1 + z_{\rm g})}}
\end{equation}

where we assumed $\Omega_{\rm m}$ = 0.272 to be consistent with the simulations described below. This allows us to normalise each 
satellite's projected group-centric distance by R$_{\rm 200}$ (i.e R$_{\rm proj}$/R$_{\rm 200}$). Combining this parameter with 
$|\Delta V|$/$\sigma$ defines the position of each satellite in the projected phase-space diagram of its host group. 
We note that Eq.(1) well reproduces the virial radius of the simulated clusters described in Sect. 3. 
As we only ever see a component of a galaxy's true velocity, or true group-centric distance, down our line-of-sight, the projected 
position has to be considered a lower limit on the true 3D position (based on the galaxy's 3D group-centric 
distance and velocity). We restrict our sample to only include satellites that fall within their host's virial 
radius in projected distance. By only considering satellites that are quite close to their host, we can expect stronger 
environmental effects. Also, in this way we can limit the contamination from interlopers (galaxies that are not associated 
with the cluster but are merely projected onto it) on our results. Beyond one virial radius, interlopers begin to dominate 
over the cluster population, but inside the zones in phase-space that we consider (see Sect. 3.1), the interloper fraction is never 
larger than 40$\%$. In fact, 92$\%$ of the group/cluster population is found in zone $\leq$ 4 where 
the fraction of interlopers is $<$ 22$\%$ and the cluster population dominates.
\par
Since sample II is not volume limited, we had to correct our basic sample for Malmquist bias, which is responsible 
for an artificial increase of the average luminosity (and M$_{\star}$) of galaxies with redshift in our data. Such an 
effect is more critical for satellites which, at fixed M$_{\rm h}$, cover a wider luminosity/mass distribution. We 
thus weighted each galaxy in our analysis by 1/V$_{\rm max}$, where V$_{\rm max}$ is the comoving volume corresponding 
to the comoving distance at which that galaxy would still have satisfied the selection criteria of the group catalogue.

\subsection{Stellar populations}
We matched our basic sample with the catalogue of stellar ages, metallicities and [$\alpha$/Fe] values of SDSS DR7 galaxies 
by Gallazzi et al. (2018 in prep). The stellar population parameters were computed following the same methodology as in 
Gallazzi et al. (2005). Briefly, the strength of the spectral indices D4000$_{\rm n}$, H$\beta$,
H$\delta_A$ $+$ H$\gamma_A$, [MgFe]$'$ and [Mg$_2$Fe] is compared with synthetic spectra built from Bruzual $\&$ Charlot
(2003) SSP models convolved with a large Monte Carlo library of star-formation histories and metallicities. This produces
probability distribution functions (PDFs) of r-band luminosity-weighted age and stellar metallicity, from which we can derive
median values and their corresponding uncertainty based on half of the (84$^{th}$ - 16$^{th}$) percentile range.
\par\noindent
For what concerns the [$\alpha$/Fe] abundance ratios, they were estimated following the approach of Gallazzi et al. (2006). 
For each galaxy, the index ratio Mgb/$<{\rm Fe}>$ (where $<{\rm Fe}>$ is the average of the Fe5270 and Fe5335
index strengths) is computed from the observed spectrum and also from
the solar-scaled model that best fits the aforementioned spectral indices (all being independent of [$\alpha$/Fe]).
The difference between data and model (denoted as $\Delta$Mgb/$<{\rm Fe}>$) can be considered as an excess of [$\alpha$/Fe] 
with respect
to the solar value. Specifically, Gallazzi et al. (2018, in prep) calculate the full PDF of $\Delta$Mgb/$<{\rm Fe}>$ from which
we then estimate a median value and its related uncertainty as half of the (84$^{th}$ - 16$^{th}$) percentile range.
Finally, Gallazzi et al. calibrate the relation between $\Delta$Mgb/$<{\rm Fe}>$  and [$\alpha$/Fe] as a function 
of age and stellar metallicity using the stellar population models of Thomas et al. (2003, 2004). It thus becomes 
possible to assign an 
[$\alpha$/Fe] value to each galaxy in the sample on the basis of its best-fit age, stellar metallicity and computed
$\Delta$Mgb/$<{\rm Fe}>$.   
\par
When matching the satellites of our basic sample with the catalogue of stellar parameters, we kept only those galaxies
with an estimated age and whose median spectral S/N per pixel is equal or larger than 20. This ensures, as shown by Gallazzi 
et al. (2005), an uncertainty on stellar age and metallicity of $\sol$ 0.2 and 0.3 dex respectively. 
The cut in S/N introduces a bias toward brighter galaxies or galaxies with deeper absorptions that may affect our 
results. In order to correct for incompleteness at fixed stellar mass we apply to each galaxy a weight $w_{SN}$ 
as computed in Gallazzi et al (2018, in prep). This weight $w_{SN}$ is calculated in bins of stellar mass (0.1 dex width) and  
absolute $(g-r)^{z=0.1}$ Petrosian colour (0.2 mag width) taken as model-independent proxy for stellar population properties. 
It is defined as the number ratio in each bin between the galaxies in the parent sample and those with spectral S/N $>$ 20. 
To avoid overweighing bins in which the statistic is low and $w_{SN}$ uncertain, we exclude bins in which high-S/N galaxies are 
fewer than 10 and less than 10$\%$ of the total. These are a negligible fraction of galaxies and does not impact our conclusions.
\par\noindent
In this way, we select 20928 satellites which make up our final working sample. 
We retrieved the median, 16$^{th}$ and 84$^{th}$ percentile values of their global, specific star formation rates (sSFR$_{\rm gl}$) 
from the catalogue by Brinchmann et al. (2004). We computed the uncertainty on the median sSFR$_{\rm gl}$ as half of the 
(84$^{th}$ - 16$^{th}$) percentile range.
\par

\begin{figure*}
\includegraphics[width=110mm]{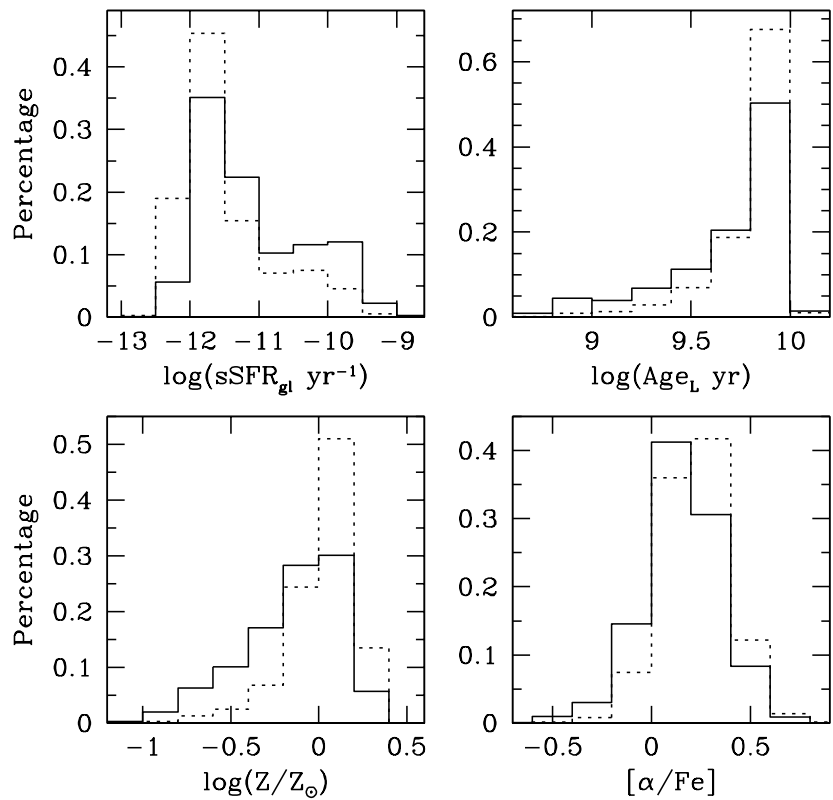}
\caption{ The weighted (solid line) and unweighted (dotted line) distribution of our working sample in the properties 
analyzed below: global specific SFR, luminosity-weighted stellar age, stellar metallicity and [$\alpha$/Fe] ratio.}
\end{figure*}

\subsection{Characterizing the working sample}
The satellites of our working sample span a redshift range between $z$ = 0.01 and $z$ = 0.2, peaking at $z \simeq$ 0.072 
and with 90$\%$ of the galaxies below $z \simeq$ 0.15. They are distributed in stellar mass between 10$^9$ and   
$\sim$10$^{11.5}$ M$_{\odot}h^{-2}$, and populate a wide range of environments, with host haloes varying in mass 
between 10$^{12}$ and 10$^{15}$ M$_{\odot}h^{-1}$. 
The solid line in Fig. 1 shows their distribution weighted by (1/V$_{\rm max}$ $\times$ $w_{SN}$) in global specific 
SFR (sSFR$_{\rm gl}$), luminosity-weighted stellar age (Age$_{\rm L}$), stellar metallicity (log (Z/Z$_{\odot}$)) and 
[$\alpha$/Fe] abundance ratio. 
About 70$\%$ of the satellites in our working sample are passive with log(sSFR$_{\rm gl}$ yr$^{-1}$) $\sol$ -11  
and luminosity-weighted stellar age older than $\sim$4 Gyr. We note that half of the working sample has solar or above solar 
metallicity, while the majority of satellites have [$\alpha$/Fe] ratios between solar and 0.5 dex. For comparison, we also trace 
in Fig. 1 the unweighted distributions of our working sample (dotted lines). They show how our weighing scheme shifts
the satellites distribution towards higher sSFR$_{\rm gl}$ and younger Age$_{\rm L}$, lower stellar metallicity and
[$\alpha$/Fe] ratio.  
\par
For a first investigation of the projected phase-space populated by our working sample, we select satellites in the stellar
mass range 10$^9$ M$_{\odot}h^{-2} \leq$ M$_{\star} \leq$ 10$^{10}$ M$_{\odot}h^{-2}$ residing in host haloes in the
mass interval 10$^{13}$ M$_{\odot}h^{-1} \leq$ M$_{\rm h} \leq$ 10$^{15}$ M$_{\odot}h^{-1}$, since Pasquali et al. (2010)
showed their properties to be most sensitive to environment. Figure 2 presents the distribution of the peculiar
velocities of these satellites, normalized by their host group velocity dispersion, as a function of their projected
distance from the luminosity-weighted centre of their group, normalized by the group virial radius R$_{\rm 200}$. 
Such distribution follows the typical trumpet shape. This arises because as galaxies with low angular momentum 
move closer to the cluster centre, they tend to gain velocity as they approach pericentre.
In order to highlight the trumpet-shape, we overlay the plot with solid lines which trace out the caustic profile 
$\Delta V / \sigma$ $\times$ R$_{\rm proj} /$R$_{\rm 200} = \pm$ 0.6. 

\begin{figure}
\includegraphics[width=85mm]{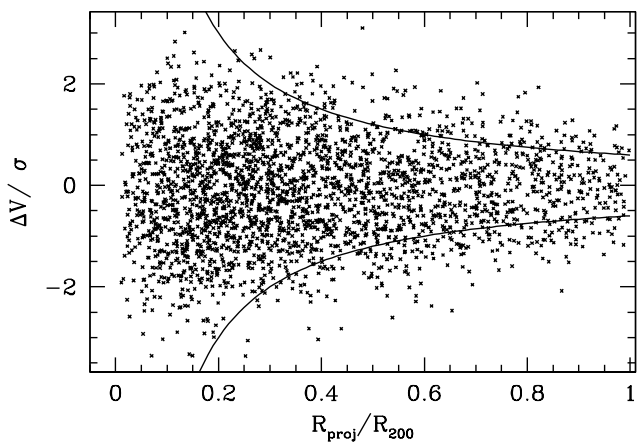}
\caption{The distribution of satellites with 10$^9$ M$_{\odot}h^{-2} \leq$ M$_{\star} \leq$ 10$^{10}$ M$_{\odot}h^{-2}$
and 10$^{13}$ M$_{\odot}h^{-1} \leq$ M$_{\rm h} \leq$ 10$^{15}$ M$_{\odot}h^{-1}$ in projected phase-space. 
The solid lines are the caustic profile $\Delta V / \sigma$ $\times$ R$_{\rm proj} /$R$_{\rm 200} = \pm$ 0.6.}
\end{figure}

In Fig. 3 we present the projected phase-space plane of Fig. 2 colour-coded on the basis of 
the fraction of passive satellites and the fraction of old and young satellites.
For each galaxy we select its 100 closest neighbours in the R$_{\rm proj}$/R$_{\rm 200}$ - $\Delta$V/$\sigma$ plane,
and separate them between passive and star-forming using the separation line derived by Oman \& Hudson (2016)
and according to their observed sSFR and stellar mass. We then compute the fraction of passive satellites, weighted 
by (1/V$_{\rm max}$ $\times$ $w_{SN}$), as a function of R$_{\rm proj} /$R$_{\rm 200}$ and $\Delta V / \sigma$ 
(top panel). We also use the luminosity-weighted stellar age to distinguish these 100 closest neighbours between
young (Age$_{\rm L} <$ 3 Gyr) and old (Age$_{\rm L} >$ 7 Gyr) and calculate their corresponding weighted fraction
(see middle and bottom panels, respectively).
Fig. 3 reveals an overall decrease in the fraction of passive or older satellites with projected cluster-centric 
distance, matched by an increase in the fraction of younger satellites. The more actively star-forming and younger
satellites are mostly located at the virial radius, but also at $|\Delta V|/ \sigma \sog$ 2. Quiescent and older 
satellites preferentially populate the inner region of their host environment. These trends with R$_{\rm proj} /$R$_{\rm 200}$
essentially occur at any fixed value of $|\Delta V|/ \sigma$. 

\begin{figure}
\includegraphics[width=110mm, angle=-90]{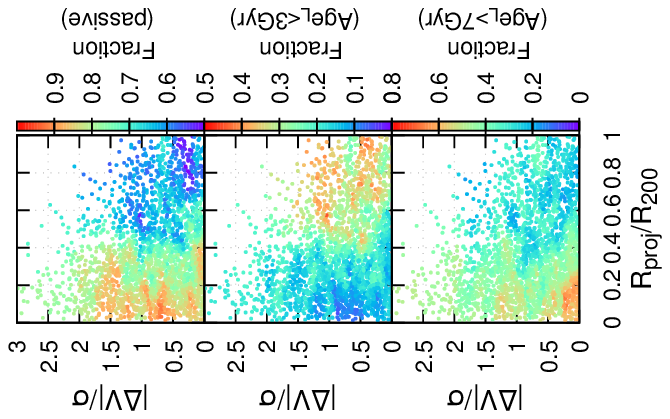}
\caption{The observed distribution of the fraction of passive satellites (top panel), the fraction of satellites
younger than 3 Gyr (middle panel) and older than 7 Gyr (bottom panel) across projected phase-space. These are computed
for satellites in our working sample in the range 10$^9$ M$_{\odot}h^{-2} \leq$ M$_{\star} \leq$ 10$^{10}$ M$_{\odot}h^{-2}$, 
residing in host haloes with 10$^{13}$ M$_{\odot}h^{-1} \leq$ M$_{\rm h} \leq$ 10$^{15}$ M$_{\odot}h^{-1}$.}
\end{figure}

\section{Galaxy cluster simulations}
In order to improve on the scatter in Fig. 3 and better understand its trends, we require some knowledge of the times at 
which our galaxies first fell 
into their current environment. For this, we must compare our observations to a set of hydrodynamic zoom-in simulations of galaxy 
clusters. Here we provide a brief summary of the hydrodynamical simulations (refereed to as YZiCS), but a full description 
is available in 
Choi $\&$ Yi (2017). We ran cosmological hydrodynamic zoom-in simulations using the adaptive mesh refinement 
code {\sc{ramses}} (Teyssier 2002). We first run a large volume cube of side length 200 Mpc~h$^{-1}$ using dark matter 
particles only, within the WMAP7 cosmology (Komatsu et al. 2011): $\Omega_{\rm{M}}$ = 0.272, 
$\Omega_{\Lambda}$ = 0.728 
and H$_0$ = 70.4 km~s$^{-1}$~Mpc$^{-1}$, $\sigma_8$ = 0.809, and n = 0.963. We then select 15 high density 
regions in the cosmological volume and perform zoom-in simulations, this time including hydrodynamic recipes. 
The zoom-in region contains all particles within 3 viral radii of a cluster at {\it z = } 0. The mass range of our 
clusters varies from 9.2$\times$10$^{14}$~M$_\odot$ to 5.3 $\times$ 10$^{13} $~M$_\odot$. A summary of their 
properties can be found in Table 1 of Rhee et al. (2017). The minimum cell size is 760 pc~h$^{-1}$ and 
8$\times$10$^{7}$~M$_\odot$ in dark matter particle mass. Halo-finding and tree-building were conducted using the 
AdaptaHOP method (Aubert et al. 2004). Each snapshot is only approximately 75 Myr apart, which greatly aids 
this process. To ensure we have a complete halo mass function, and to reduce numerical effects, we discard all 
subhaloes whose peak mass is lower than 3 $\times$10$^{10}$~M$_{\odot}$ from our analysis. With this mass cut, all haloes can be 
considered as a proxy for a galaxy, and, even in the worst case, we can follow halo mass loss down to 3$\%$ of 
the peak mass. We have adopted the baryon prescriptions of Dubois et al. (2012) including gas cooling, star 
formation, stellar and AGN feedback, and were able to reproduce the basic properties of observed $z \sim$0 galaxies
to the same degree as in the Horizon-AGN simulations (Dubois et al. 2014). 
\par\noindent
In order to compare our simulations with the data, we compute a projected phase-space diagram for each individual 
cluster, considering 1000 line of sights (increasing this number by a factor of five introduces negligible 
differences, cf. Rhee et al. 2017). We then combine all the cluster phase-space planes together so that each of 
them has equal weight in the final projected phase-space in order to avoid biases due to varying numbers of cluster members. 
\par
To this set of simulations we add data extracted from a suite of 8 zoom-in cosmological simulations of clusters. These 
simulations are fully described in Warnick $\&$ Knebe (2006). In short, these are N-body simulations that were performed 
with the adaptive mesh refinement code {\sc mlpam} (Knebe, Green $\&$ Binney 2001). The simulations are dark-matter only but the 
resolution is roughly comparable with the mass and spatial resolution of the YZICS simulations. Each dark matter particle has 
a mass of $\sim$1.6$\times$10$^8$~M$_\odot$, the highest spatial resolution is $\sim$2~kpc, and the 8 clusters are in the mass 
range 1--3$\times$10$^{14}$~M$_\odot$. Smith et al. (2015) systematically measured the orbital parameters (eccentricity and 
pericentre distance of the most recent orbit) for all surviving subhaloes within these cluster simulations (see blue shading in 
Fig. 10 of Smith et al. 2015). For convenience, we use this existing data set to look at 
how these orbital parameters vary across phase-space, stacking together all clusters into a single phase-space 
diagram. We exclude all objects which are yet to reach a first pericentre passage as their orbital parameters are not yet defined.

\begin{figure*}
\includegraphics[width=170mm]{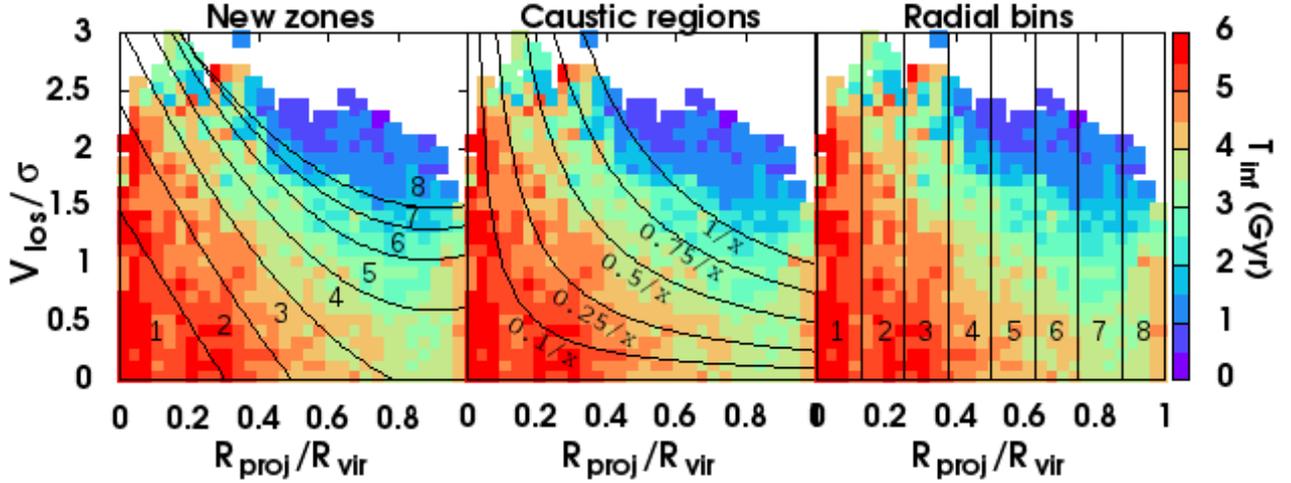}
\caption{The distribution of infall time, T$_{\rm inf}$, in Gyr across the projected 
phase-space, computed by averaging the simulated galaxies in 2D bins. {\it Left-hand panel:} The family of 
curves defined by Eq. (3) and Eq. (4) is plotted with solid black lines, and the zones in which these 
curves split the projected phase-space are numbered. {\it Middle panel:} several representative caustic
profiles, which follow the analytical expression $y = c/x$ with $y = |\Delta V|/ \sigma$
and $x$ = R$_{\rm proj}$/R$_{\rm vir}$. {\it Right-hand panel:} the projected phase-space is
dissected among 8 equally wide radial bins.} 
\end{figure*}

\subsection{Discretizing phase-space} 

Caustic profiles (as shown in Fig. 2) have usually been used in the literature to separate kinematically
different galaxy populations in phase-space, which were accreted onto their parent halo at different times 
(e.g. Mamon et al. 2004; Gill et al. 2005; Mahajan et al. 2011; Haines et al. 2012). In particular, Noble et al. 
(2013, 2016) introduced plots of galaxy observed properties as a function of caustic profile defined as 
$|\Delta V|$/$\sigma$ $\times$ R$_{\rm proj}$/R$_{\rm 200}$, as different values of this parameter generally correspond
to different infall times. 
\par\noindent
We check how closely caustic profiles trace the accretion history of a massive halo
by computing the mean infall time, T$_{\rm inf}$, in two-dimensional bins across the projected phase-space 
of the satellites in the hydrodynamic simulations set. T$_{\rm inf}$ is defined as the time since a 
galaxy crossed for the first time the virial radius of the main progenitor of its present-day host environment. 
The result is shown in Fig. 4, where the distribution of simulated satellites in phase-space is colour-coded on the basis 
of their mean T$_{\rm inf}$ (Gyr) computed in 2D bins. Galaxies that have not yet entered 
the cluster are excluded from this plot as they do not have a T$_{\rm inf}$ value, and thus interlopers are also excluded. 
As already pointed out by Rhee et al. (2017), T$_{\rm inf}$ decreases with
both increasing R$_{\rm proj}$/R$_{\rm vir}$ and $|\Delta V|$/$\sigma$, so that the most recently accreted 
($\sim$1.5 Gyr ago) galaxies have a relatively high peculiar velocity and are found in the outskirts of their host halo, 
while the satellites that fell in earliest ($\sim$5.5 Gyr ago) typically have a smaller peculiar velocity and a smaller, 
projected cluster-centric distance. The comparison of Fig. 4 with Fig. 3 highlights a correspondence between
the distribution of satellite fractions and the distribution of T$_{\rm inf}$ across projected phase-space. Also,
the vertical stripes of constant fraction value in Fig. 3 are reminiscent of the zones of constant T$_{\rm inf}$ in Fig. 4,
although they appear less pronounced, possibly because of the contribution from interlopers that are 
excluded in Fig. 4. Such a similarity points to more quiescent satellites with the stellar populations that appear old today 
having been accreted earlier onto their present-day host halo. 
\par\noindent
In the middle panel of Fig. 4 we over-plot a few representative caustic profiles, whose analytical form is
$y = c/x$, where $y = |\Delta V| / \sigma$, $x$ = R$_{\rm proj}$/R$_{\rm vir}$ and $c$ a numerical coefficient
which varies across the projected phase-space. In particular, we define the following set of caustic profiles:
\begin{equation}
\begin{split}
1: y & = 0.10/x \\ 
2: y & = 0.25/x \\
3: y & = 0.50/x \\
4: y & = 0.75/x \\
5: y & = 1.00/x \\
6: y & > 1.00/x
\end{split}
\end{equation}
\par\noindent
We also split the projected phase-space in 8 equal width radial bins in the right-hand side panel of Fig. 4 (we do so 
because it is common use in the literature to show how satellite properties vary as a function of cluster-centric distance). 
It is clear that both caustic profiles and radial bins do not well match the shape of the infall time distribution. 
We thus replace them with a family of quadratic curves which are customised to split up the projected phase-space into eight new 
zones within 1 R$_{\rm vir}$, which follow the infall time distribution, as plotted 
in the left-hand side panel of Fig. 4. Each curve has the analytical form: 

\begin{equation}
|\Delta V| / \sigma = a{\rm (R_{\rm proj}/R_{\rm vir})}^2 + b{\rm (R_{\rm proj}/R_{\rm vir})} + c 
\end{equation}

\par\noindent
where the coefficients $a$, $b$ and $c$ are expressed as a function of $p$, an integer number running between 1 and 7: 

\begin{equation}
\begin{split}
a & = 0.022p^3 - 0.512p^2 + 3.287p - 2.786 \\ 
b & = 0.184p^2 - 1.494p - 3.5 \\
c & =-0.108p^2 + 1.249p + 0.314
\end{split}
\end{equation}

\par\noindent
For example, $p$ = 1 gives the dividing line between zone 1 and zone 2.
In Fig. 5 we present the distribution in projected phase-space of the number density of the simulated satellites. 
The curves defined by Eq. (3) and Eq. (4) are overlaid and the new zones numbered. We can see that the  average
number density decreases with increasing zone number as expected, but is still high to statistically justify the
definition of zones $\geq$ 5. 

\begin{figure}
\includegraphics[width=90mm]{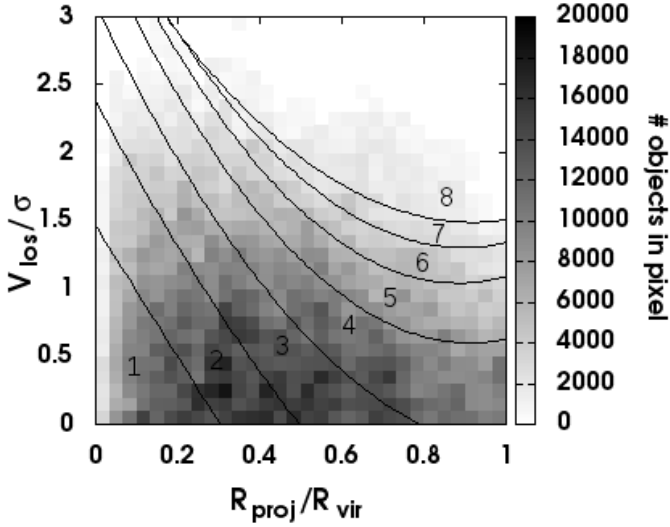}
\caption{The distribution of the number density of the simulated galaxies  
across projected phase-space. The curves defined by Eq. (3) and Eq. (4) are overplotted as black lines,
and the zones which they separate are labelled with a zone number.}
\end{figure}

\begin{table}
\centering
\caption{The average infall time $\overline{\rm T}$$_{\rm inf}$ and its associated standard deviation 
($\sigma$($\overline{\rm T}$$_{\rm inf}$)) as derived for the 8 new zones, the 6 
caustic regions as well as the 8 equally wide radial bins drawn in Fig. 4. We also list the first and
third quartiles of the T$_{\rm inf}$ distribution within each of these areas.} 
\begin{tabular}{ccccccccccccccc}
\hline
Zones & First & $\overline{\rm T}$$_{\rm inf}$ & Third &$\sigma$($\overline{\rm T}$$_{\rm inf}$) \\ 
   & quartile &        & quartile &     \\ 
   & (Gyr)    & (Gyr)  &  (Gyr)   & (Gyr)\\
\hline
1 & 3.41 & 5.42 & 7.13 & 2.51 \\ 
2 & 3.02 & 5.18 & 7.13 & 2.60 \\
3 & 2.33 & 4.50 & 6.39 & 2.57 \\
4 & 2.10 & 3.89 & 5.66 & 2.34 \\
5 & 1.58 & 3.36 & 5.08 & 2.36 \\
6 & 0.99 & 2.77 & 4.20 & 2.29 \\
7 & 0.78 & 2.24 & 2.71 & 1.97 \\
8 & 0.49 & 1.42 & 1.80 & 1.49 \\
\hline
Caustic & First & $\overline{\rm T}$$_{\rm inf}$ & Third & $\sigma$($\overline{\rm T}$$_{\rm inf}$)\\ 
regions  & quartile &        & quartile &       \\ 
         & (Gyr)    & (Gyr)  &  (Gyr)   & (Gyr)\\
\hline
1 & 3.02 & 5.08 & 6.89 & 2.53 \\
2 & 2.55 & 4.68 & 6.56 & 2.55 \\
3 & 2.10 & 4.22 & 6.23 & 2.52 \\
4 & 1.50 & 3.49 & 5.49 & 2.43 \\
5 & 1.06 & 2.84 & 4.36 & 2.31 \\
6 & 0.56 & 1.76 & 2.33 & 1.66 \\
\hline
Radial & First & $\overline{\rm T}$$_{\rm inf}$ & Third & $\sigma$($\overline{\rm T}$$_{\rm inf}$) \\
bins  & quartile &        & quartile &  \\
      & (Gyr)    & (Gyr)  &  (Gyr)   & (Gyr)\\
\hline
1 & 3.41 & 5.38 & 6.97 & 2.57 \\
2 & 2.63 & 4.95 & 6.97 & 2.67 \\
3 & 2.40 & 4.90 & 6.89 & 2.71 \\
4 & 1.87 & 4.21 & 6.31 & 2.63 \\
5 & 1.65 & 3.77 & 5.74 & 2.54 \\
6 & 1.73 & 3.53 & 5.49 & 2.38 \\
7 & 1.80 & 3.10 & 4.36 & 2.06 \\
8 & 2.33 & 3.54 & 4.60 & 2.02 \\
\hline
\end{tabular}
\end{table}

Table 1 lists the average T$_{\rm inf}$ ($\overline{\rm T}$$_{\rm inf}$), its standard deviation 
($\sigma$($\overline{\rm T}$$_{\rm inf}$)), and the first and third quartiles of the T$_{\rm inf}$
distribution as computed for the 8 new zones, the 6 caustic regions and the 8 equal width radial bins drawn 
in Fig. 4. Note that the values of $\sigma$($\overline{\rm T}$$_{\rm inf})$ in the zones is consistent
with those obtained for the 2D bins used in Fig. 4. 
Although the different regions can not be directly
compared as they cover different areas of the projected phase-space, we can see that, in most cases, 
$\sigma$($\overline{\rm T}$$_{\rm inf}$) is reduced with the zones, especially when compared with high
number radial bins. In addition, the range of $\overline{\rm T}$$_{\rm inf}$ is larger for the zones (4 Gyr)
than for the caustic regions (3.3 Gyr) and the radial bins (2 Gyr). 
A larger range of $\overline{\rm T}$$_{\rm inf}$ is crucial, as it means that we are more sensitive to changes 
in the infall time of the galaxy population by using our zones than with the other methods. In summary, the 
comparison with the radial bins shows that, by additionally considering cluster-centric velocities, we can 
better distinguish between galaxy populations with differing mean infall time, and this is accomplished 
better using our zones approach than with the caustic method.

\begin{table*}
\centering
\caption{The mean infall time $\overline{\rm T}$$_{\rm inf}$ (and its standard deviation $\sigma$($\overline{\rm T}$$_{\rm inf}$)
in Gyr), the average fraction of stripped dark matter subhalo $<$ f$_{\rm dm} >$ (with its standard deviation 
$\sigma$($<$ f$_{\rm dm} >$)), the fraction of interlopers f$_{\rm intl}$ and the fraction of simulated galaxies which
have not yet experienced a pericentric passage f$_{\rm no-per}$ as a function of zone in the YZiCS simulation of Choi \& Yi (2017).} 
\begin{tabular}{ccccccc}
\hline
Zone & $\overline{\rm T}$$_{\rm inf}$ & $\sigma$($\overline{\rm T}$$_{\rm inf}$) &
$<$ f$_{\rm dm} >$ & $\sigma$($<$ f$_{\rm dm} >$) & f$_{\rm intl}$ & f$_{\rm no-per}$\\
\hline
1 & 5.42 & 2.51 & 0.73 & 0.18 & 0.06 & 0.13\\
2 & 5.18 & 2.60 & 0.67 & 0.19 & 0.07 & 0.18\\
3 & 4.50 & 2.57 & 0.61 & 0.21 & 0.10 & 0.23\\
4 & 3.89 & 2.34 & 0.53 & 0.22 & 0.13 & 0.29\\
5 & 3.36 & 2.36 & 0.49 & 0.23 & 0.17 & 0.39\\
6 & 2.77 & 2.29 & 0.46 & 0.24 & 0.22 & 0.48\\
7 & 2.24 & 1.97 & 0.40 & 0.23 & 0.26 & 0.54\\
8 & 1.42 & 1.49 & 0.40 & 0.22 & 0.40 & 0.60\\
\hline
\end{tabular}
\end{table*}

Since our aim is to investigate any possible dependence of galaxy properties on projected phase-space for satellites of
different M$_{\star}$ and living in different environments, we need to check whether the trend of decreasing 
T$_{\rm inf}$ with increasing zone number, as seen in Fig. 4, is also recovered 
if we control for the effects of stellar and halo mass. 
For this purpose we have separated the simulated sample into two host 
halo mass bins (13 $<$ log(M$_{\rm h}) <$ 14 and 14 $<$ log(M$_{\rm h}) <$ 15) and the simulated satellites into 
two bins in stellar
mass (9 $<$ log(M$_{\star}) <$ 10 and log(M$_{\star}) >$  10). For simplicity, we calculate the stellar masses from the 
satellites halo mass based on the halo abundance matching prescription of Guo et al. (2010). We plot the mean infall 
time (solid lines) as a function of zone for each mass bin and for the general satellite population in Fig. 6. Here, 
we also show the first and third quartiles of the T$_{\rm inf}$ distribution within each zone (dashed lines). 
As we can see, the separation into different halo/stellar mass bins does not significantly affect the trend already 
highlighted in Fig. 4 for the general population of satellites: all trends in $\overline{\rm T}$$_{\rm inf}$, as well
as in the 25$^{th}$ and 75$^{th}$ percentiles, exhibit the same gradient, indicating that the zones drawn in Fig. 4 
consistently work well at distinguishing more recent infalls from infalls that happened a long time ago. Therefore,
Figs. 4 and 5 confirm that our definition of zones in projected phase-space tracks the stronger change in {\it mean} 
infall time as we move from zone 1 to zone 8. We note that a small shift towards shorter $\overline{\rm T}$$_{\rm inf}$ with 
increasing satellite stellar mass exists, and an even smaller shift with increasing host mass exists, although in both 
cases, it is very minor compared to the broad spread in $\overline{\rm T}$$_{\rm inf}$ in each zone. These shifts are
a natural consequence of hierarchical accretion as also discussed in De Lucia et al. (2012).  
\par\noindent
Although our simulations do not allow us to check whether the relation between zones and $\overline{\rm T}$$_{\rm inf}$
holds also in less massive hosts, we will later apply our zone scheme also to observed satellites
residing in environments less massive than 10$^{13}$ M$_{\odot}$.

\begin{figure}
\includegraphics[width=80mm]{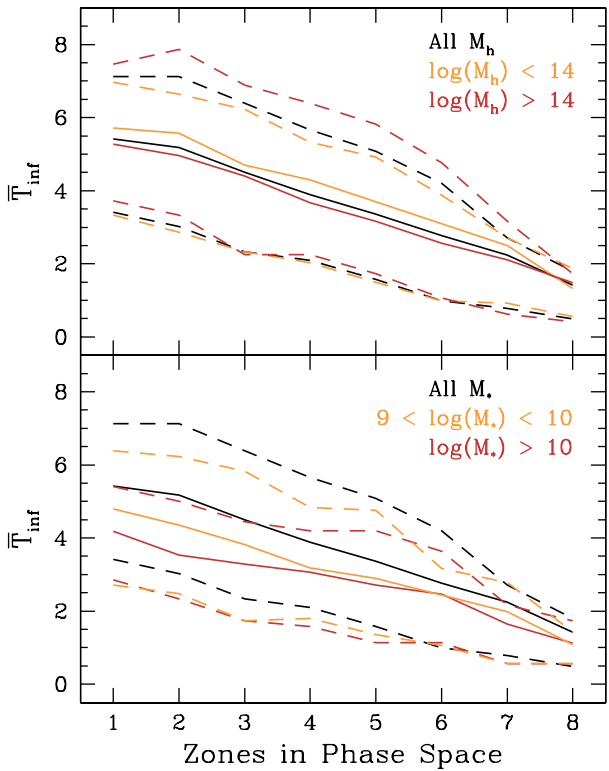}
\caption{Solid lines trace the mean infall time, $\overline{\rm T}$$_{\rm inf}$, in Gyr as a function of zone in 
projected phase-space and in different bins of halo mass (top panel) and stellar mass (bottom panel) as derived from the 
YZiCS hydrodynamic simulations. The 25$^{th}$ and 75$^{th}$ percentiles of the T$_{\rm inf}$ distribution within each zone
are plotted with dashed lines. The simulated dwarf galaxies are not shown in this figure because the log(M$_{\star}) < $ 9 
bin is not included in the analysis of our data; for each zone, their $\overline{\rm T}$$_{\rm inf}$ is slightly longer
than that of the `All M$_{\star}$' sample.}
\end{figure}

Table 2 summarizes the main properties of the YZiCS simulation set in each zone in projected phase-space, including 
$\overline{\rm T}$$_{\rm inf}$ in Gyr, and its associated standard deviation $\sigma$($\overline{\rm T}$$_{\rm inf}$), 
as well as the average fraction of dark matter subhalo stripped from satellites $<$ f$_{\rm dm} >$ and its standard deviation 
$\sigma$($<$ f$_{\rm dm} >$). We also report in the table the fraction of interlopers f$_{\rm intl}$\footnote{in this case, 
defined as objects beyond 3 virial radius that are simply projected into the phase-space diagram.} 
and the fraction of satellites (interlopers excluded) that have not yet had a pericentric passage f$_{\rm no-per}$.
In Appendix A we discuss in detail how much projection effects result in cross-contaminations of the zones in terms
of T$_{\rm inf}$. 
\par
As these results are derived from one set of cosmological simulations, it is likely that we can expect some changes 
in $\overline{\rm T}$$_{\rm inf}$ and shape of the T$_{\rm inf}$ distribution in each zone if using a different set of 
simulations. These differences could arise from having different sample size, differing cosmological parameters, or more 
technical issues such as different mass and spatial resolution, or differing definitions of haloes and their boundaries. 
Nevertheless, we are confident that {\it the mean T$_{\rm inf}$ will always systematically decrease with increasing zone number}. 
Indeed we have confirmed that this is the case in the Millennium simulations. This fact is key to our methodology and 
the main conclusions of this study as it enables us to see how the galaxy population properties change when we alter 
the population's average T$_{\rm inf}$.

\section{Satellite observed properties in projected phase-space}
Numerical and hydro-dynamical simulations of environmental effects such as ram-pressure and tidal stripping show that
satellites are deprived of their gas and stars outside-in, and the amplitude of this removal should depend on their
stellar mass, their orbit within their host halo, the dark-matter mass of the host, and the time when they were accreted 
onto their host environment. As a consequence, we expect to detect a dependence of the observed properties of satellites
in their projected phase-space since this is simply a parameterization of their orbits and hence infall histories.  

\begin{figure*}
\includegraphics[width=120mm]{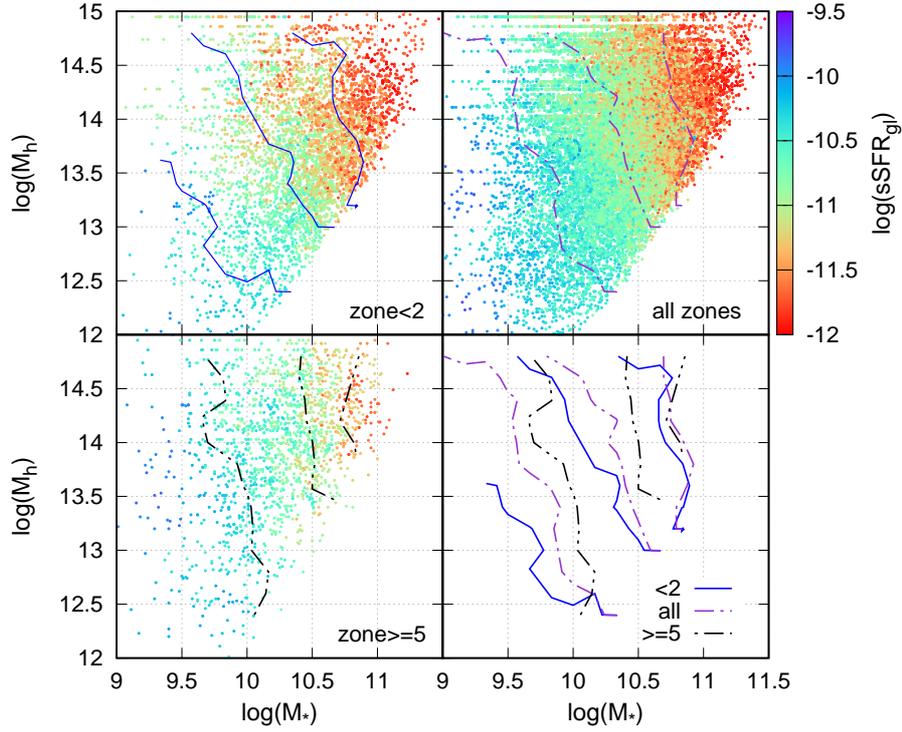}
\caption{Galaxy specific star formation rate as a function of galaxy M$_{\star}$  and host M$_{\rm h}$
for our working sample. Contours are shown at
log(sSFR$_{gl}$) = -10.5, -11.0, -11.5 (from left to right). The upper-right panel shows the full
sample. The upper-left panel shows galaxies in zone $\leq$ 2 ($\overline{\rm T}$$_{\rm inf} \geq$ 5 Gyr).
The lower-left panel shows galaxies in zone $\geq$ 5 ($\overline{\rm T}$$_{\rm inf} \leq$ 3.4 Gyr). The lower-right
panel shows how the contour shape and position change when $\overline{\rm T}$$_{\rm inf}$ shifted.}
\end{figure*}

\begin{figure*}
\includegraphics[width=120mm]{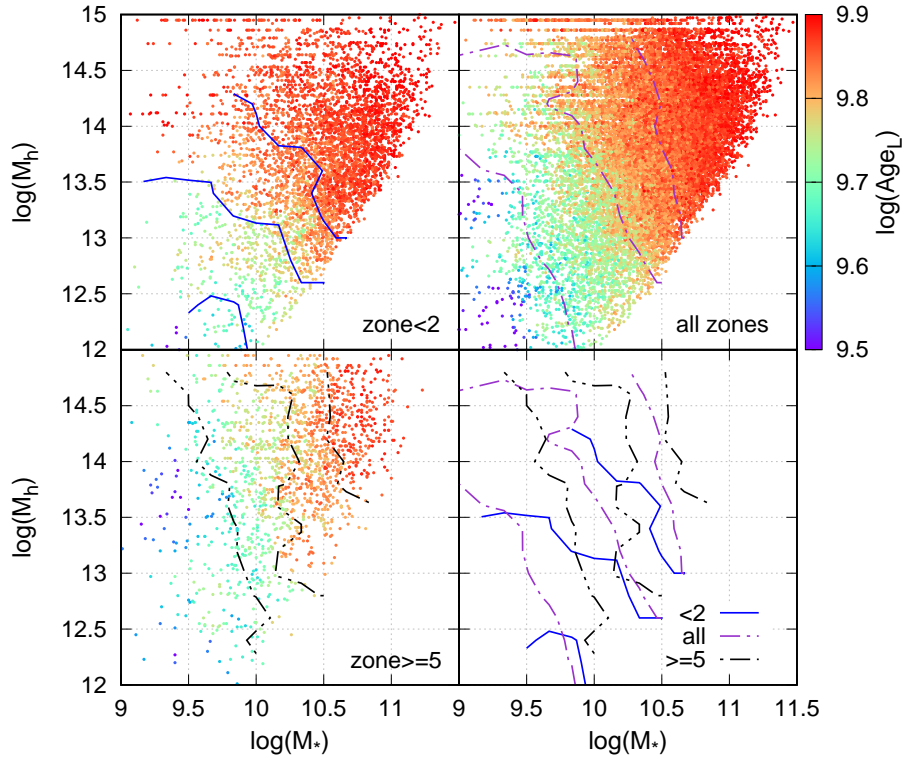}
\caption{As in Fig. 6 but for luminosity-weighted age instead of specific star formation rate. Contours
are shown at log(Age$_L$) = 9.7, 9.8, 9.85 (from left to right). Galaxies in zone $\leq$ 2 have 
$\overline{\rm T}$$_{\rm inf} \geq$ 5 Gyr, while those in zone $\geq$ 5 have
$\overline{\rm T}$$_{\rm inf} \leq$ 3.4 Gyr.}
\end{figure*}

\par

\begin{figure*}
\includegraphics[width=120mm]{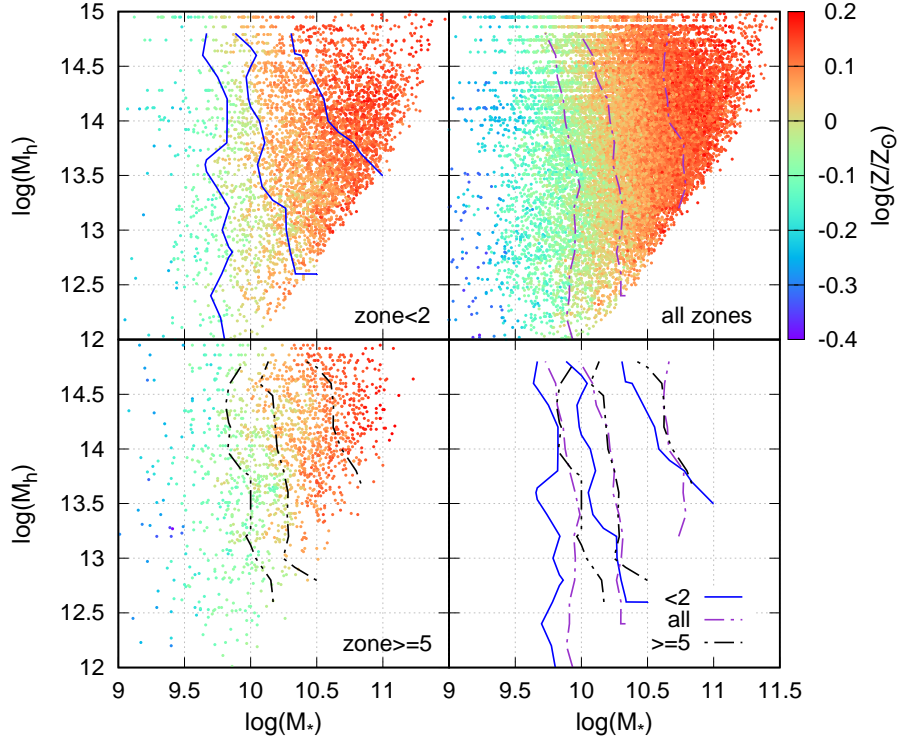}
\caption{As in Fig. 6 but for stellar metallicity instead of specific star formation rate. Contours
are shown at log(Z/Z$_\odot$) = -0.025, 0.05, 0.125 (from left to right). Galaxies in zone $\leq$ 2 have  
$\overline{\rm T}$$_{\rm inf} \geq$ 5 Gyr, while those in zone $\geq$ 5 have
$\overline{\rm T}$$_{\rm inf} \leq$ 3.4 Gyr.}
\end{figure*}

\begin{figure*}
\includegraphics[width=120mm]{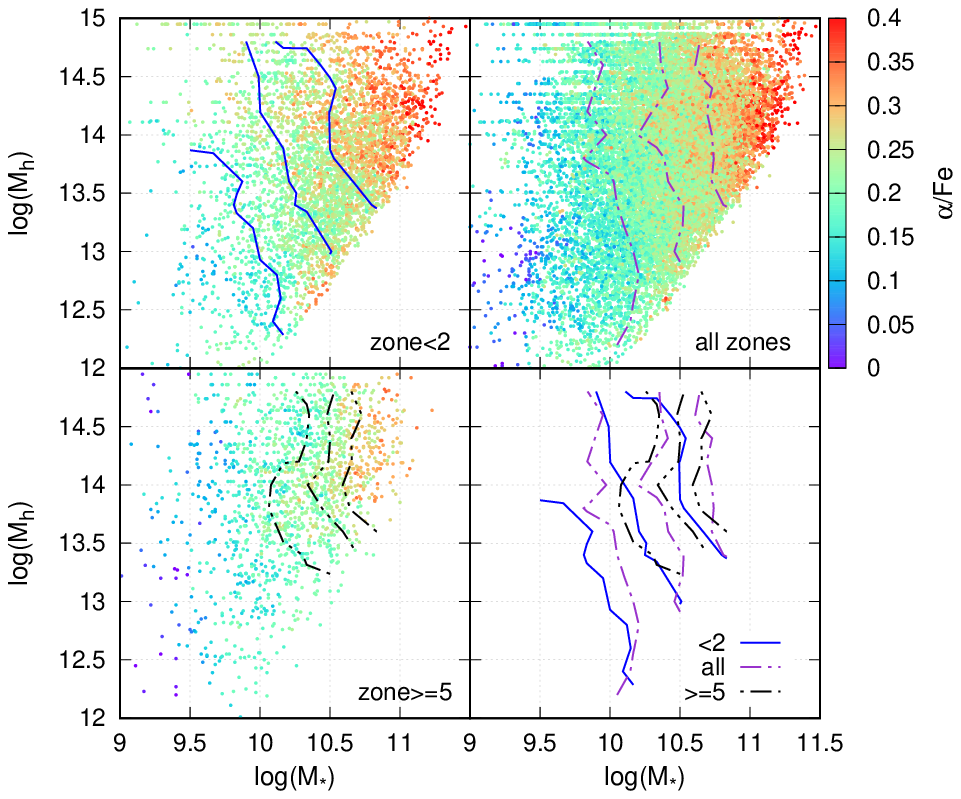}
\caption{As in Fig. 6 but for [$\alpha$/Fe] element ratio instead of specific star formation rate.
Contours are shown at [$\alpha$/Fe] = 0.2, 0.24, 0.275 (from left to right). Galaxies in zone $\leq$ 2 have  
$\overline{\rm T}$$_{\rm inf} \geq$ 5 Gyr, while those in zone $\geq$ 5 have
$\overline{\rm T}$$_{\rm inf} \leq$ 3.4 Gyr.}
\end{figure*}

\subsection{Observed global distributions}
We use the $|\Delta V|$/$\sigma$ and R$_{\rm proj}$/R$_{\rm 200}$ values measured for our working sample
and Eq.(3) and (4) to assign each satellite a zone in projected phase-space and thus a $\overline{\rm T}$$_{\rm inf}$ value.
We then plot each galaxy's location in the log(M$_{\star}$) and log(M$_{\rm h}$)
plane, and colour the point by galaxy property. In this way, we can effectively see how galaxy properties vary as we
simultaneously control for the effect of galaxy mass and host mass (which is our proxy for
environment)\footnote{These plots can be
considered analogous to Fig. 6 from Peng et al. (2010).}. In order to look for large scale trends across the plot,
we plot for each galaxy the average property of its nearest 30 neighbours in the log(M$_{\star}$) - log(M$_{\rm h}$)
plane.
This average also takes into account the weighing required
to correct for our observational biases (discussed in Sect. 2). The results are shown in Figs. 7 to 10. In each plot we
consider a different galaxy property, and in each panel we split the sample by zone. Satellites with zone $<$ 2
(corresponding to $\overline{\rm T}$$_{\rm inf} \geq$ 5 Gyr) and zone $\geq$ 5 (with $\overline{\rm T}$$_{\rm inf} \leq$ 3.4 Gyr)
are shown in the top left and bottom left panels, respectively, and all zones combined are
shown in the top right panel.
\par\noindent
We overlay contours at three fixed
values on each distribution, and compare them in the bottom right panel of Figs. 7 - 10.
To make the contours, we first bin up the log(M$_{\rm h}$)-log(M$_{\star}$) plane presented in the Figures into a 15 $\times$ 15 pixels
grid, and compute the mean value in each pixel. Contours are then fitted to the gridded values.
\par\noindent
A visual inspection of the all-zone distributions clearly shows that, at fixed M$_{\rm h}$,
satellites become more passive and older, as well as metal-richer and enriched in $\alpha$-elements, with increasing
M$_{\star}$. A weaker trend in host halo mass is detectable for sSFR$_{\rm gl}$ and Age$_{\rm L}$ at fixed M$_{\star}$
(cf. Pasquali et al. 2010), which disappears in the distributions of recent infallers, i.e. satellites with zone $\geq$ 5.
The picture though changes for the ancient infallers in zone $<$ 2, since their sSFR$_{\rm gl}$, Age$_{\rm L}$ and possibly
[$\alpha$/Fe] appear to depend on M$_{\star}$ and more clearly on halo mass. In fact, at fixed M$_{\star}$, sample satellites grow
more passive and older with increasing M$_{\rm h}$ (this dependence is weaker for [$\alpha$/Fe]). These changes
already start to occur at the lower host masses, a feature that does not emerge in the all-zone distributions: {\it
we first need to control for the effect of infall time in order to clearly bring out the effects of lower mass hosts.}
\par\noindent
At the same time, we see that the contours progressively move to lower stellar and halo masses as we shift zone from 5 to 2,
and hence increase $\overline{\rm T}$$_{\rm inf}$ at fixed contour level. This is clearly seen for sSFR$_{\rm gl}$ and Age$_{\rm L}$,
and, at lower extent, for stellar metallicity and [$\alpha$/Fe].
We also note how, at fixed contour level, these contours become less steep and bend over with increasing
$\overline{\rm T}$$_{\rm inf}$, particularly in the case of sSFR$_{\rm gl}$ and Age$_{\rm L}$.
\par\noindent
{\it These results indicate that
sSFR$_{\rm gl}$ and Age$_{\rm L}$ of a satellite primarily depend on its stellar mass when such satellite is a recent
infaller, while it additionally depends on M$_{\rm h}$ when the satellite was accreted at earlier times, i.e.
when environmental effects have been at play for a longer time. Moreover, the dependence
of sSFR$_{\rm gl}$ and Age$_{\rm L}$ of ancient infallers on M$_{\rm h}$ indicates that, upon
infall, the environment takes several Gyrs to quench satellites, and more massive haloes are more efficient in extinguishing
the star formation activity of satellites.} In the case of stellar metallicity and [$\alpha$/Fe], the separation of our sample satellites
into different zones (or differing $\overline{\rm T}$$_{\rm inf}$) does not highlight any significant trend of log(Z/Z$_{\odot}$)
with M$_{\rm h}$ at fixed M$_{\star}$. In other words, the galaxies seem to obey the mass-metallicity relation, nearly
independent of their environment.
\par\noindent
One caveat is that, in Figure 6, we can see that more massive satellites (log(M$_{\star}$ M$_{\odot}h^{-2}) >$ 10) have slightly
reduced $\overline{\rm T}$$_{\rm inf}$ values than less massive satellites (9 $<$ log(M$_{\star}$ M$_{\odot}h^{-2}) <$ 10),
in particular at the zone 1 end,
although the difference is small ($<$0.75~Gyr) and well within the scatter in the trend. For Figs. 7 to 11, this implies there is a
small decrease in $\overline{\rm T}$$_{\rm inf}$ along the log(M$_{\star}$) axis, affecting the zone $<$ 2 panels only. This small
decrease in $\overline{\rm T}$$_{\rm inf}$ might slightly reduce environmental quenching, and thus the effect might
be to fractionally counteract the mass quenching trend in the zone $<$ 2 panels only.

\subsection{Trends of satellite observed properties with zone}
We now turn to investigate in more detail the dependence of the satellite properties on zone in projected phase-space, 
using a more quantified and less visual approach than previously. For this purpose, we split the satellites of our working 
sample in 2D bins defined by stellar mass and halo mass as follows:
\begin{itemize}
\item {9 $\leq$ log(M$_{\star}$ M$_{\odot}h^{-2}) <$ 10, since Pasquali et al. (2010) showed that satellites in this 
stellar mass range are most sensitive to environmental effects;}
\item {10 $\leq$ log(M$_{\star}$ M$_{\odot}h^{-2}) <$ 10.5, which we consider a transition range, representing galaxies
which are less affected by environmental processes;}
\item {10.5 $\leq$ log(M$_{\star}$ M$_{\odot}h^{-2}) \leq$ 11.5, since Pasquali et al. (2010) found that the stellar
properties of such massive satellites are the least environmentally dependent;} 
\end{itemize}
\par\noindent
and:
\begin{itemize}
\item {12 $\leq$ log(M$_{\rm h}$ M$_{\odot}h^{-1}) <$ 13, the low mass environments;}
\item {13 $\leq$ log(M$_{\rm h}$ M$_{\odot}h^{-1}) <$ 14, the intermediate mass host haloes;}
\item {14 $\leq$ log(M$_{\rm h}$ M$_{\odot}h^{-1}) \leq$ 15, the massive environments.}
\end{itemize}
\par\noindent 
This stacking is meant not only to provide sufficient statistics for our analysis, but also to reduce the noise 
due to errors on the observed line-of-sight velocity and cluster-centric distance of the satellites in our
working sample. The galaxy statistics available for these 2D bins is summarized in Table 3.
For each 2D mass bin, we derive the average of a specific galaxy property weighted by (1/V$_{\rm max} \times w_{SN}$)
from all satellites belonging to the same zone in phase-space, and also compute 
the standard error on the weighted mean. We consider only those zones which contain at least 20 satellites each. 

\begin{figure*}
\includegraphics[width=180mm]{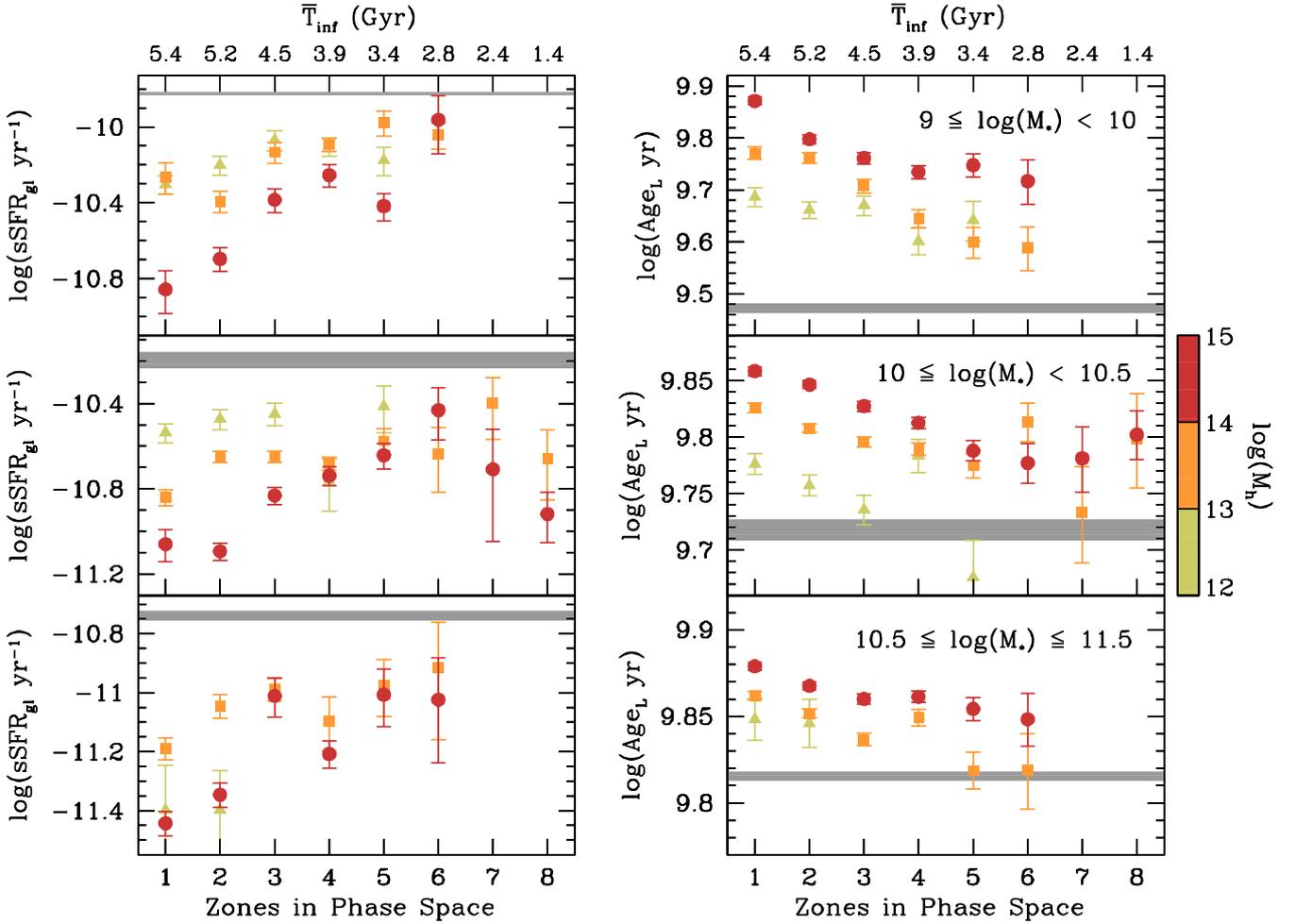}
\caption{The observed distribution of average global sSFR and average age of our sample satellites as a function
of zone and $\overline{\rm T}$$_{\rm inf}$ in projected phase-space
in the left-hand and right-hand column respectively. The error bars represent the error on the mean.
Satellites are split into three bins of stellar mass (from 10$^{9}$ to 10$^{11.5}$ M$_{\sun}h^{-2}$) and 3 bins of
halo mass: 12 $\leq$ log[M$_{\rm h}$ (M$_{\sun}h^{-1}$)] $<$ 13 (yellow triangles), 13 $\leq$ log[M$_{\rm h}$
(M$_{\sun}h^{-1}$)] $<$ 14 (orange squares) and 14 $\leq$ log[M$_{\rm h}$ (M$_{\sun}h^{-1}$)] $\leq$ 15
(red circles). Each grey shaded area represents the range in weighted mean sSFR$_{\rm gl}$ and Age$_{\rm L}$
spanned by field galaxies.}
\end{figure*}

\begin{table}
       \centering
       \caption{The number of satellites in our working sample per bin of  log(M$_{\star}$~M$_{\odot}h^{-2}$) and
log(M$_{\rm h}$~M$_{\odot}h^{-1}$).}
       \begin{tabular}{c|cccc}
                   \hline
                & &  & log(M$_{\rm h}$) & \\
            log(M$_{\star}$)  & & 12 - 13 & 13 - 14 & 14 - 15\\
               \hline
                9.0 - 10.0 & & 775 &1267 &1257\\
               10.0 - 10.5 & & 921 &4468 &3377\\
               10.5 - 11.5 & & 74 &4176 &4428\\
               \hline
       \end{tabular}
\end{table}

\subsubsection{Specific SFR as a function of \/ $\overline{\rm T}$$_{\rm inf}$}
In the left-hand column of Fig. 11 we plot the average global sSFR as a function of  zone and $\overline{\rm T}$$_{\rm inf}$ 
for satellites separated into the three M$_{\star}$ bins and colour-coded on the basis of the three M$_{\rm h}$ bins 
defined above. Each grey shaded area shows the range in  weighted mean sSFR$_{\rm gl}$ 
spanned by field galaxies in each M$_{\star}$ bin (see Sect. 4.3).
{\it A trend of increasing log(sSFR$_{\rm gl}$) with increasing zone number, hence decreasing 
$\overline{\rm T}$$_{\rm inf}$, is visible between zones 1 to 6, which is stronger for satellites in  
more massive haloes (red circles).} We estimate by how much sSFR$_{\rm gl}$ varies between zones 1 and 6,  
and see that this variation $\Delta$ increases with halo mass.
At log(M$_{\star}$ M$_{\odot}h^{-2}) \leq$ 10.5, $\Delta$ is $\sim$0.15, $\sim$0.25 and $\sim$0.65 dex 
in low-, intermediate- and high-mass hosts, respectively. For 
the more massive satellites, $\Delta$ is $\sim$0.3 and 0.4 dex in intermediate- and high-mass haloes, respectively. 
An interesting feature of the intermediate-mass satellites in $>$10$^{13}$ M$_{\odot}h^{-1}$ haloes is the
flattening of their distribution to lower sSFR$_{\rm gl}$ values at zone $>$ 6. 
\par\noindent
We also notice that, for satellites less massive than log(M$_{\star}$ M$_{\sun}h^{-2}$) = 10.5 and in zones $<$ 4, 
the average sSFR$_{\rm gl}$ is generally higher in low-mass haloes, and decreases with increasing M$_{\rm h}$. {\it In other words, 
at fixed stellar mass and fixed zone (for the same $\overline{\rm T}$$_{\rm inf}$),
the quenching of the star-formation activity of satellites is controlled by the halo mass of the host, and 
is more efficient in more massive hosts.} 
\par
The above trends are qualitatively consistent with the results of Hern\'andez-Fern\'andez et al. (2014) who found that 
colour-selected passive galaxies in 16 clusters at $z <$ 0.05 are preferentially distributed in phase-space within the
caustic profile  ($| \Delta {\rm V}|/ \sigma_{\rm los})$ $\times$ (R$_{\rm p}$/R$_{\rm 200}) \sol $ 0.4 (corresponding
to our zones $\sol$ 3 in Fig. 4), while the distribution of star-forming galaxies extends to higher caustic profiles 
(or higher zone numbers). 
Similarly to our results for $z \sim$ 0 galaxy groups and clusters, Noble at al. (2013) and Noble et al. (2016) observed 
a remarkable increase ($\sim$1 - 2 dex) in sSFR with increasing values of 
($| \Delta {\rm V}|/ \sigma_{\rm los})$ $\times$ (R$_{\rm p}$/R$_{\rm 200}$), hence with increasing zone number, in a 
cluster at $z$ = 0.87 and in three clusters at $z$ = 1.2. More recently, Barsanti et al. (2018) found that the fraction
of star-forming galaxies in the Galaxy And Mass Assembly (GAMA) group catalogue (at 0.05 $\leq z \leq$ 0.2) is higher in 
the outer region of their projected phase-space (corresponding to higher zone numbers).

\begin{figure*}
\includegraphics[width=180mm]{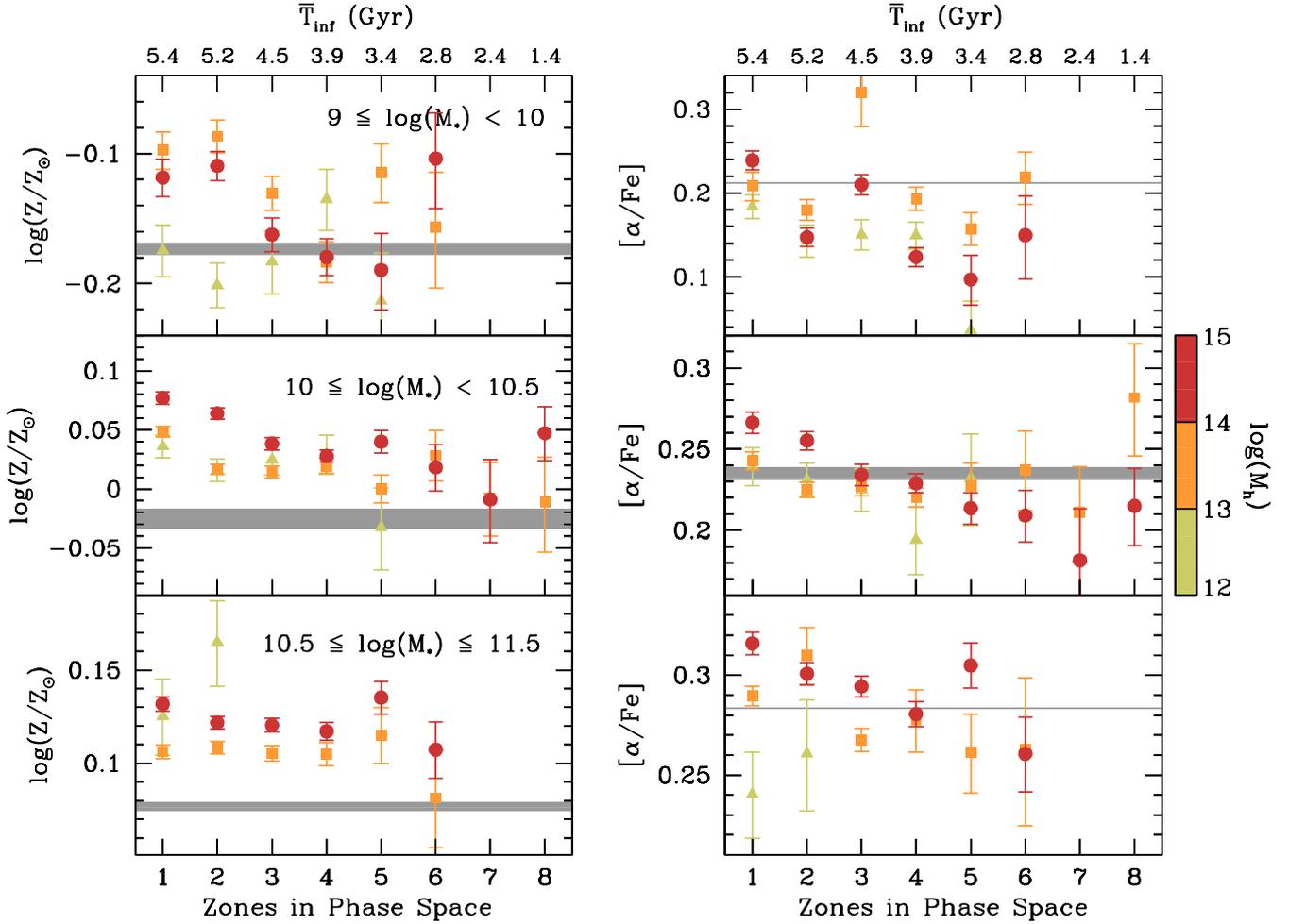}
\caption{As in Fig. 11 but now for average stellar metallicity and average [$\alpha$/Fe] ratio of satellites
in our working sample. Each grey shaded area represents the range in weighted mean log(Z/Z$_{\odot}$) and [$\alpha$/Fe]
spanned by field galaxies.}
\end{figure*}

\subsubsection{Stellar age as a function of \/ $\overline{\rm T}$$_{\rm inf}$}
The right-hand column of Fig. 11 shows the average, luminosity-weighted stellar age of satellites as a function of 
zone and $\overline{\rm T}$$_{\rm inf}$ for the same M$_{\star}$ and M$_{\rm h}$ bins as 
in the left-hand column. Also in this case the grey stripes show the ranges in Age$_{\rm L}$ covered by field galaxies. 
\par\noindent
{\it The behaviour 
of stellar age is opposite to that of average sSFR$_{\rm gl}$, as it clearly decreases with increasing zone number 
(thus with decreasing $\overline{\rm T}$$_{\rm inf}$), at least up to zone 6.}
The amplitude $\Delta$ of the age variation between zone 1 and 6 is larger for low mass satellites 
(M$_{\star} <$ 10$^{10}$ M$_{\sun}h^{-2}$), and it increases from $\sim$0.7 Gyr in low-mass host haloes (yellow
triangles) to $\sim$2 Gyr in host haloes more massive than log(M$_{\rm h}$ M$_{\sun}h^{-1}$) = 13 (orange
squares and red circles). For intermediate-mass satellites $\Delta$ is $\sim$1 Gyr, while it decreases to
$\sim$0.6 Gyr at M$_{\star} >$ 10$^{10.5}$ M$_{\sun}h^{-2}$. 
As in the case of sSFR$_{\rm gl}$, we point to the flattening of the age distribution of intermediate-mass 
satellites in host haloes more massive than 10$^{13}$ M$_{\sun}h^{-1}$ at zone $>$ 6.
\par\noindent
{\it We also observe that, at fixed stellar mass and zone (for zone $<$ 6), satellites become progressively older as the 
halo mass of their host groups increases, which adds further support for the M$_{\rm h}$-driven quenching of
galaxy star formation.} At fixed host halo mass the mean stellar age of satellites can be seen to increase with stellar 
mass (see also Pasquali et al. 2010). 
\par
The above results are qualitatively similar to the analysis performed by Noble et al. (2013) and Noble et al. (2016), who 
traced the dependence of the age indicator D4000 on caustic profiles for galaxies in clusters at $z$ = 0.87 and 1.2.
They showed that D4000 steadily decreases with increasing ($| \Delta {\rm V}|/ \sigma_{\rm los})$ $\times$ 
(R$_{\rm p}$/R$_{\rm 200})$ (hence with increasing zone number), so that galaxies exhibit progressively younger ages 
for larger values of caustic profile.

\subsubsection{Stellar metallicity as a function of \/ $\overline{\rm T}$$_{\rm inf}$}
We plot the average stellar metallicity as a function of zone and $\overline{\rm T}$$_{\rm inf}$, and per bin
of stellar and halo mass in the left-hand column of Fig. 12 (the 
grey shaded areas represent the range in weighted mean log(Z/Z$_{\odot}$)  
of field galaxies). {\it Despite the noise in the distributions, 
a weak trend of decreasing metallicity with increasing zone number (hence decreasing $\overline{\rm T}$$_{\rm inf}$) can be 
recognised at stellar masses log(M$_{\star}$ M$_{\sun}h^{-2}$) $\leq$ 10.5 and log(M$_{\rm h}$ M$_{\sun}h^{-1}$) $\geq$ 13.} 
Its amplitude between zones 1 and 6  
is $\Delta \sim$ 0.05 dex for low mass satellites, and $\Delta \sim$ 0.04 dex for intermediate-mass
satellites. The average stellar metallicity of more massive satellites is essentially invariant with zone.
The Z distribution of intermediate-mass satellites in massive haloes shows a flattening at zone $>$ 6,
where stellar metallicity levels off within the error bars. 
\par\noindent
In the case of satellites more massive than M$_{\star}$ = 10$^{10}$ M$_{\sun}h^{-2}$ and for zone $<$ 5 the
average log(Z/Z$_{\odot}$) slightly increases with M$_{\rm h}$ at fixed stellar mass and zone (similarly to 
that found by Pasquali et al. 2010, although less pronounced).
We also notice that, at fixed halo mass, the average log(Z/Z$_{\odot}$) increases with stellar mass as 
expected from the known galaxy metallicity - mass relation (e.g. Tremonti et al. 2004, Gallazzi et al. 2005,
Mannucci et al. 2010, Foster et al. 2012, S\'anchez et al. 2017).

\subsubsection{[$\alpha$/Fe] ratio as a function of \/ $\overline{\rm T}$$_{\rm inf}$}
The [$\alpha$/Fe] element ratio is shown as a function of zone and $\overline{\rm T}$$_{\rm inf}$
in the right-hand column of Fig. 12 for satellite and field galaxies.
{\it First of all, we notice that, regardless of zone, at fixed halo mass, [$\alpha$/Fe] increases 
with stellar mass, but no systematic dependence of [$\alpha$/Fe] on halo mass at fixed M$_{\star}$ can be found.} 
This reproduces the findings of Gallazzi et al. (2018, in prep) that the relation between [$\alpha$/Fe] and M$_{\star}$ 
is to first order independent of halo mass. 
{\it Here, by studying the distribution of [$\alpha$/Fe] in projected phase-space we can identify a trend of decreasing [$\alpha$/Fe] 
with increasing zone number (decreasing $\overline{\rm T}$$_{\rm inf}$) in particular for satellites in haloes 
more massive than 10$^{14}$ M$_{\odot}h^{-1}$.} For this halo mass range, the amplitude $\Delta$ of the overall change in 
[$\alpha$/Fe] between zones 1 and zones 6 depends on M$_{\star}$: it varies from $\sim$0.1 dex at low stellar masses to 
$\sim$0.05 dex for intermediate and high masses.
\vskip 0.2truecm
\par\noindent
We have checked that the above trends do not significantly depend on our weighing scheme.  
In the extreme case when galaxy properties are weighted only by 1/V$_{\rm max}$, the trends in Figs. 11 and 12 are seen
to shift to older ages, higher metallicity and [$\alpha$/Fe] values, and to lower sSFR$_{\rm gl}$ values by an amount typically 
less than 3$\sigma$, while their gradients become less steep. In order to check the effect of the uncertainty on  
the group centre definition, we have computed mean galaxy properties as a function
of zone using the projected distance of satellites from their central galaxy, and weighing galaxy properties by 
(1/V$_{\rm max} \times w_{SN}$). The new trends are largely consistent with those shown in Figs. 11 and 12.
We have also calculated how the mean observed galaxy properties change as a function of the caustic regions
defined in Sect. 3.1, where $|\Delta V|$/$\sigma$ $\times$ R$_{\rm proj}$/R$_{\rm 200}$ varies from 0.1 to 1.0 in steps
of 0.1. We find trends similar to those obtained with our zones approach, albeit more noisy.

\subsection{Comparison with the field}
It is interesting to check how the trends of the observed properties of satellites in projected phase-space compare 
with the average properties of field galaxies.
\par\noindent
We use the DR7 group catalogue of Wang et al. (2014) to define the field population using three different criteria:
{\it (i)} by extracting all central galaxies residing in haloes less massive than M$_{\rm h}$ = 10$^{12}$ 
M$_{\sun}h^{-1}$; {\it (ii)} by retrieving all centrals with no associated satellite (down to the magnitude
limit of the SDSS spectroscopy); {\it (iii)} by selecting all galaxies of {\it (ii)} which are more than 5R$_{\rm 200}$
away from their closest environment which is not their host halo. In all cases, we keep only those galaxies 
whose spectral S/N is 20 or
higher. We notice that criteria {\it (i)} provides us with centrals less massive than M$_{\star}$ = 10$^{10.5}$ M$_{\sun}h^{-2}$,
while the centrals chosen by criteria {\it (ii)} and {\it (iii)} populate the whole stellar mass range of the satellites 
in our working sample. 
\par\noindent 
For each sample, we compute the weighted average sSFR$_{\rm gl}$, Age$_{\rm L}$, log(Z/Z$_{\odot}$) and
[$\alpha$/Fe] in each stellar mass bin. We then plot the range defined by the mean galaxy properties of 
the three samples as a grey shaded area in Figs. 11 and 12.
\par
Field galaxies are more strongly star forming and have younger luminosity-weighted ages at any M$_{\star}$ 
than equally massive satellites, although in some cases satellites in zone $\geq$ 6 are 
consistent with the field population at the 1 - 2$\sigma$ level. The average age of low mass satellites exhibits the 
strongest deviation from the field, indicating once again that they are the galaxies most prone to environmental effects. 
\par\noindent
We observe a clear offset in log(Z/Z$_{\odot}$) between the field population and satellites more massive than log(M$_{\star}$ 
M$_{\sun}h^{-2}$) = 10, in that satellites in zone $<$ 6 are metal-richer by $\sim$0.05 dex. Satellites with higher zone 
number have a stellar metallicity more similar to that of field galaxies. In the case of low mass satellites, we see a 
trend with halo mass: those residing in low mass haloes are as metal-rich as field galaxies at the 1$\sigma$ level, while 
satellites living in intermediate- and high mass haloes are $\sim$0.06 dex metal-richer than the field if in zone $<$ 3, else  
have log(Z/Z$_{\odot}$) similar to that of field galaxies at the 1 - 2$\sigma$ level. 
\par\noindent
Finally, only satellites with 
log(M$_{\star}$ M$_{\odot}h^{-2}) >$ 10, log(M$_{\rm h}$ M$_{\odot}h^{-1}) >$ 14 and zone $<$ 3 show a [$\alpha$/Fe] abundance 
ratio slightly higher than the field, while they become comparable with the field population at zone $>$ 3 similarly to 
equally massive satellites in lower mass haloes at any zone. Low mass satellites in low- and high mass environments display 
[$\alpha$/Fe] values lower than the field. 
\par
{\it In conclusion, we see satellite properties generally approaching the field values with increasing zone number,
and nearly equalling the field for low mass hosts. On the contrary, the average observed properties of satellites in more
massive environments often seem to flatten out before reaching the field values.}

\section{Discussion}
Dissecting the projected phase-space of galaxy groups/clusters into zones of constant mean infall time
has allowed us to emphasize the dependence of galaxy properties on environment while controlling for the effects 
of both galaxy and halo mass. In what follows we will discuss the environmental processes responsible for the
observed trends in projected phase-space, 
we will also use the observed dependence of galaxy properties on $\overline{\rm T}$$_{\rm inf}$ to estimate
time-scales for the quenching of star formation, and to address satellite pre-processing. 
We will refer to satellites in low number zones as 'ancient infallers', and to satellites
in high number zones as 'recent infallers'.

\subsection{Nurture at work}
The information on zones in projected phase-space has allowed us to bring out the dependence of
sSFR$_{\rm gl}$ and Age$_{\rm L}$ of ancient infallers on halo mass (cf. Figs. 7 and 8, also Barsanti et al. 2018),
which would otherwise go undetected when mixing satellites of different T$_{\rm inf}$ (see Peng et al. 2010).
More specifically, at fixed M$_{\star}$ the satellites' sSFR$_{\rm gl}$ smoothly increases with zone while 
their Age$_{\rm L}$ decreases, indicating that ancient infallers are more quenched than recent infallers.
In addition, at fixed zone, satellites' sSFR$_{\rm gl}$ and Age$_{\rm L}$ tend to decrease and increase, respectively,
with increasing halo mass (cf. Fig. 11). We interpret these trends as the result of the early T$_{\rm inf}$ of
ancient infallers, which thus underwent gas loss via strangulation and ram-pressure stripping at earlier times and
for longer durations. The strength of these environmental effects is obviously amplified in more massive host haloes
because of their deeper potential well and denser intracluster medium, both of which make gas removal and star-formation
quenching more efficient.
\par
The trends in the distribution of metallicity as a function of zone are far less pronounced than those in
sSFR$_{\rm gl}$ and Age$_{\rm L}$. However, a general increase of Z with decreasing zone can
be recognized for satellites less massive than log(M$_{\star}$ M$_{\odot}h^{-2}) =$ 10.5 and
residing in more massive haloes. These results are likely the effect of ram-pressure stripping with some
possible contribution from tidal stripping. As ram-pressure stripping proceeds outside-in, it removes the
satellite outskirts that are typically metal-poor, and inhibits radial gas inflows which otherwise would dilute the
gas-phase metallicity in the central region of the satellite (see also Pasquali et al. 2012). Bah\'e et al. (2017)
showed that the absence of metal-poor gas inflows allows a satellite to form stars from metal-rich gas and hence
to increase its overall stellar metallicity by an amount which seems to be consistent with the metallicity difference
between satellites and the field population in Fig. 12. The longer exposure to ram pressure stripping
can thus justify the higher metal content of ancient infallers with respect to recent ones. 
\par\noindent
This scenario is somehow corroborated by the fact that satellites have similar [$\alpha$/Fe] as the
field population, indicating a similar duration of their star formation activity although it occurs at lower
sSFR$_{\rm gl}$ in satellites. Therefore, the higher metal content of satellites with respect
to the field is not due to stronger or more efficient star formation in satellites, but possibly to star formation
taking place in metal-richer gas. We note here that the weak trend of increasing [$\alpha$/Fe] with $\overline{\rm T}$$_{\rm inf}$
possibly indicates an early truncation of star formation in ancient infallers, when [$\alpha$/Fe] is
used as a proxy for the duration of the star formation activity in galaxies (cf. de La Rosa et al. 2011). Nevertheless,
the galaxy formation models of De Lucia et al. (2017) have highlighted that [$\alpha$/Fe] may not be a straightforward
indicator of the time-scale of star formation as it significantly depends on the intensity of star formation and
on other galaxy properties such as AGN feedback (cf. Segers et al. 2016), galactic winds and the Initial Mass Function
(cf. Fontanot et al. 2017). 
\par\noindent
The trend of increasing log(Z/Z$_{\odot}$) with $\overline{\rm T}$$_{\rm inf}$ could also be due to tidal stripping.
If a galaxy initially sits on the normal stellar mass-metallicity relation, then after losing stellar mass
it may appear too metal-rich for its new stellar mass, compared to typical galaxies (see also
Pasquali et al. 2010). Increasing the time spent in a host might result in a larger stellar mass loss, which would
thus explain the trend of rising stellar metallicity with increasing $\overline{\rm T}$$_{\rm inf}$.
In the following, we attempt to roughly estimate how much stellar mass would need
to be stripped to explain the average metallicity of zone 1 satellites. First, we define the relation
between stellar mass and metallicity followed by all centrals which have a spectral S/N $\ge$ 20 in the Gallazzi et al.'s
(2018 in prep) catologue, by calculating the weighted average metallicity in small bins of stellar mass.
We then use this relation to read the stellar mass corresponding to the average metallicity of zone 1 satellites,
and we dub the difference between this value and the mean M$_{\star}$ of zone 1 satellites as stellar mass loss.
We find that such stellar mass loss is typically 0.25 dex, quite independently of the stellar and halo mass
as also found by Han et al. (2018) using the same YZiCS simulations.
On average, 80$\%$ stellar mass loss would be necessary to explain the metallicity of ancient infallers through
tidal stripping of their progenitor.
\par\noindent
This value is somehow higher than the fraction of stripped dark matter (f$_{\rm dm}$ in Table 2) in zone 1, and, together
with the fact that dark-matter haloes are more easy to strip, suggests that stellar mass loss alone is unlikely to fully
explain our result of increasing log(Z/Z$_{\odot}$) with $\overline{\rm T}$$_{\rm inf}$.
A fraction of the satellites in zone 1 may have suffered enough tidal mass loss to reach these metallicities, but
they are not enough to match the mean observed metallicity. 
\par\noindent

\begin{table*}
\centering
\caption{The mean pericentric distance ($<$R$_{\rm per}$/R$_{\rm vir}>$) and orbit eccentricity ($< e >$),
computed in projected phase-space from the simulations of Warnick \& Knebe (2006), as a function
of zone and mean infall time $\overline{\rm T}$$_{\rm inf}$ (in Gyr).}
\begin{tabular}{cccccc}
\hline
Zone & $\overline{\rm T}$$_{\rm inf}$ & $<$R$_{\rm per}$/R$_{\rm vir}>$ & $\sigma$($<$R$_{\rm per}$/R$_{\rm vir}>$) & $< e >$ &
$\sigma$($< e >$)\\
\hline
1 & 5.42 & 0.220 &  0.233 & 0.552 & 0.221 \\
2 & 5.18 & 0.296 &  0.280 & 0.559 & 0.232 \\
3 & 4.50 & 0.355 &  0.319 & 0.574 & 0.243 \\
4 & 3.89 & 0.436 &  0.392 & 0.585 & 0.253 \\
5 & 3.36 & 0.516 &  0.429 & 0.557 & 0.266 \\
6 & 2.77 & 0.547 &  0.403 & 0.552 & 0.277 \\
7 & 2.24 & 0.581 &  0.414 & 0.555 & 0.281 \\
8 & 1.42 & 0.588 &  0.437 & 0.562 & 0.286 \\
\hline
\end{tabular}
\end{table*}

\begin{figure*}
\vskip 0.3truecm
\includegraphics[width=150mm]{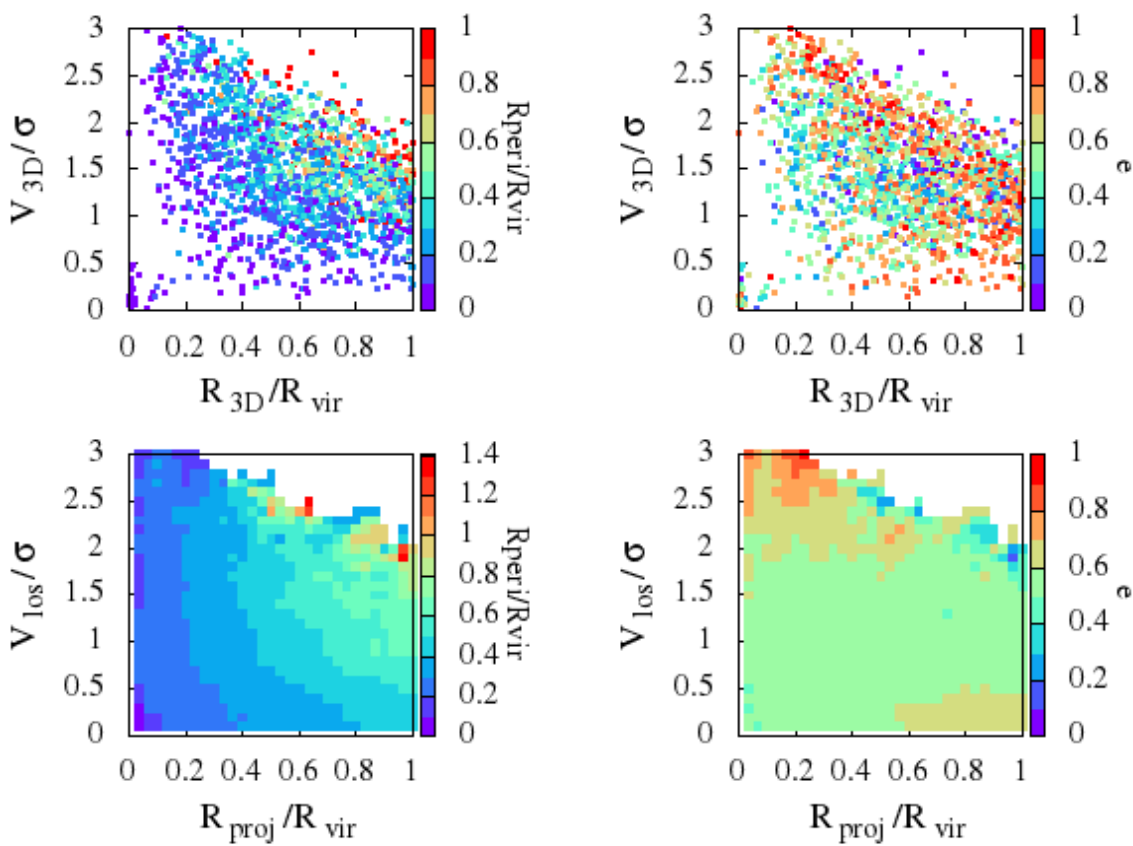}
\caption{{\it Top panels:} the distributions of pericentric distance in units of virial radius, R$_{\rm peri}$/R$_{\rm vir}$,
and orbit eccentricity, $e$, across the 3D phase-space diagram in Warnick \& Knebe's (2006) simulations. {\it Bottom panels:} 
the distributions of R$_{\rm peri}$/R$_{\rm vir}$ and $e$ on the 2D, projected phase-space diagram in Warnick \& Knebe's (2006)
simulations.}
\end{figure*}

\subsubsection{The effect of orbital parameters}
Ram-pressure and tidal stripping not only depend on the dark matter density distribution of a halo,
or the density of its intracluster medium, but also on the orbital velocity of a galaxy in that halo.
All of these should vary with location along the orbit, and the variation might be expected to be larger for objects
on more eccentric orbits. Figure 1 of Rhee et al. (2017) presents a toy representation of a typical trajectory that an
infalling galaxy takes on a 3D phase-space diagram. Although both orbital velocity and cluster-centric radius fluctuate
with time, overall objects tend to shift toward lower orbital velocity and cluster-centric distance as time
passes by (see also the left panel of Figure 2 of Rhee et al. 2017). This change is partly the result of dynamical friction,
which steadily dissipates the orbital energy of satellites, especially for massive galaxies. In addition, all satellites
are influenced by the deepening and changing shape of the potential well in which they orbit as their
hosts haloes grow in mass by accretion. This effect is also thought to be responsible for satellite orbits becoming
increasingly circularised with time (Gill et al. 2004). 
The net result of these effects is that, with increasing T$_{\rm inf}$, satellites are found closer to the
central region of their host halo, where the intracluster medium is denser and the halo potential well is deeper.
If their orbits become more circularised, they will spend longer periods of time near the centre, and may become
increasingly quenched and tidally stripped (of gas and stars). It is thus interesting to use the simulations to compute
the mean orbital parameters, pericentric distance and orbit eccentricity, as a function of zone and
$\overline{\rm T}$$_{\rm inf}$ (see Table 4), and compare them with the trends of observed galaxy properties in 
projected phase-space.
\par
In the top panels of Fig. 13 we plot the satellites simulated by Warnick \& Knebe (2006) in a 3D phase-space diagram. 
The data points are colour
coded on the basis of their normalised pericentre distance R$_{\rm peri}$/R$_{\rm vir}$ (left-hand panel), and their
orbital eccentricity $e$ (right-hand panel). In a 3D phase-space diagram, it is clear that both pericentre distance and
eccentricity vary accross the diagrams. For example, we see that the pericentric distance increases with cluster-centric
distance and also with orbital velocity. The distribution is somewhat similar to the distribution of
T$_{\rm inf}$ in projected phase-space, as shown in Fig. 4. The orbital eccentricity distribution is a
little more complex with high values of eccentricity found at both high and low velocities simultaneously. Objects with a
combination of both small pericentric distances and small orbit eccentricities (i.e. rounder orbits) tend to be located in
more central regions of the 3D phase-space diagram.

In the bottom panels of Fig. 13, we present projected phase-space diagrams. For these we consider 100 randomly chosen
line-of-sights, and build up the mean normalised pericentre distance (lower-left panel) and orbital eccentricity distribution
(lower-right panel) in each pixel. The projected distribution of pericentre distance has a similar trend to the 3D version,
only it appears more smoothed. However, projection effects wash out most of the trend seen in the 3D orbital eccentricity plot.
These results are also confirmed quantitatively by the mean values of R$_{\rm peri}$/R$_{\rm vir}$ and $e$ shown in Table 4,
which we computed for each zone, together with their standard deviation. Ancient infallers are,
on average, characterized by small pericentric distances (R$_{\rm peri}$/R$_{\rm vir}$ $\simeq$ 0.2 - 0.3), while recent
infallers have  R$_{\rm peri}$/R$_{\rm vir}$ $\sim$ 0.6. We conclude that the large T$_{\rm inf}$ of ancient infallers has
resulted in a population of galaxies with small pericentric distance and lowered eccentricity, thus making these satellites
spend more time in the central region of their host haloes. This could partly contribute to the fact that we see the strongest
environmental effects in this population, as these galaxies will spend most of the time near their host's centre where environmental
effects are likely most effective. However, if the impact of some environmental mechanisms is accumulated with time, then environmental
effects would be most visible in the population of galaxies that have spent most time in the cluster, and thus the change in orbital
parameters may only provide an additional contribution to the overall change. We also note that, although we can confirm
that orbital circularisation can be seen to occur in our 3D phase space diagram, a lot of the eccentricy variation is washed out
in the projected phase space diagrams, and as a result it may be difficult to detect orbital circularisation in observed phase
space diagrams.

\subsection{Further implications of the observed trends in project phase-space}
\subsubsection{Quenching time-scales}

\begin{table*}
\centering
\caption{The gradient and intercept of the linear fits of observed sSFR$_{\rm gl}$ {\it vs.}
$\overline{\rm T}$$_{\rm inf}$, together with the estimated quenching time-scale $\Delta t_{\rm q}$
and their corresponding errors.}
\begin{tabular}{cccccc}
\hline
log(M$_{\star}$ M$_{\odot}h^{-2}$) & Parameters &  Units &      & log(M$_{\rm h}$ M$_{\odot}h^{-1}$)  & \\
                                    &            &        & 12 - 13 & 13 - 14 & 14 - 15\\
\hline
9.0 - 10.0 & $\frac{\Delta(\rm {sSFR_{\rm gl}})}{\Delta t}$ & 10$^{-11}$yr$^{-1}$Gyr$^{-1}$ & -2.398 $\pm$ 0.185
& -2.370 $\pm$ 0.159 & -3.197 $\pm$ 0.229\\
           & Intercept & 10$^{-11}$yr$^{-1}$                & 15.834 $\pm$ 0.249 & 15.854 $\pm$ 0.226
& 15.863 $\pm$ 0.251 \\
           & $\Delta t_{\rm q}$ & Gyr                       &  6.187 $\pm$ 0.488 &  6.268 $\pm$ 0.432
&  4.649 $\pm$ 0.342 \\
           &                                                &   &     &        &        \\
10.0 - 10.5  & $\frac{\Delta(\rm {sSFR_{\rm gl}})}{\Delta t}$ & 10$^{-11}$yr$^{-1}$Gyr$^{-1}$ & -0.861 $\pm$ 0.189
& -1.151 $\pm$ 0.114 & -1.273 $\pm$ 0.047\\
             & Intercept & 10$^{-11}$yr$^{-1}$                &  6.232 $\pm$ 0.509 &  6.211 $\pm$ 0.268
&  6.327 $\pm$ 0.106 \\
             & $\Delta t_{\rm q}$ & Gyr                       &  6.075 $\pm$ 1.456 &  4.528 $\pm$ 0.504
&  4.183 $\pm$ 0.176 \\
           &                                                &   &    &        &        \\
10.5 - 11.5  & $\frac{\Delta(\rm {sSFR_{\rm gl}})}{\Delta t}$ & 10$^{-11}$yr$^{-1}$Gyr$^{-1}$ & -0.309 $\pm$ 0.009
& -0.223 $\pm$ 0.054 & -0.283 $\pm$ 0.067\\
             & Intercept & 10$^{-11}$yr$^{-1}$                &  1.734 $\pm$ 0.033 &  1.724 $\pm$ 0.191
&  1.723 $\pm$ 0.225\\
             & $\Delta t_{\rm q}$ & Gyr                       &  2.375 $\pm$ 0.129 &  3.243 $\pm$ 1.164
&  2.551 $\pm$ 0.997\\
\hline
\end{tabular}
\end{table*}

As the phase-space analysis provides us with constraints on the average time that galaxies 
have spent in their hosts, we can now examine how their observed mean sSFR$_{\rm gl}$ changes 
as a function of time in their host, for hosts and satellites of different mass. To do so, we 
consider zones 1 to 6, and associate them with their $\overline{\rm T}$$_{\rm inf}$ as listed 
in Table 2. We can also use the field sSFR$_{\rm gl}$ value as our $\overline{\rm T}$$_{\rm inf} =$ 0 
data point. This is the mean among the field sSFR$_{\rm gl}$ values derived with methods $(i)$, $(ii)$
and $(iii)$ described in Sect.4.3, and its associated error is half the full sSFR$_{\rm gl}$ range
computed for field galaxies at fixed stellar mass. 
In general, we find that the time evolution of the observed decline in mean sSFR$_{\rm gl}$  
has a very linear form. Only in zones 7 and 8 of the intermediate mass satellites do we find some 
deviation from the linear fit, where some reduced values of sSFR$_{\rm gl}$ are found (probably due 
to pre-processing as discussed below), and so we neglect these data points when conducting the 
linear fit. We then perform a simple linear fit of observed mean sSFR$_{\rm gl}$ {\it vs.} 
$\overline{\rm T}$$_{\rm inf}$, which takes into account the errors on sSFR$_{\rm gl}$ (as shown in Fig. 11)
and the standard deviation on $\overline{\rm T}$$_{\rm inf}$ (as listed in Table 2).  
The resulting best-fit gradient, intercept and associated errors are reported in Table 5 together with their units. 
\par\noindent
In general, comparing between columns in the table, we can see that the rate of reduction of the 
mean sSFR$_{\rm gl}$ after entering a host is higher for more massive hosts, possibly because of their
denser intracluster medium. 
This trend is though not visible in our highest 
stellar mass bin, thus suggesting that internal mass quenching is more dominant for these galaxies,
which, in addition, tend to be accreted later. 
In any case, even for the massive satellites, their sSFR$_{\rm gl}$  continues to decline with time spent 
in their host, albeit at a slower rate than for less massive galaxies. 
Comparing between stellar masses (i.e. comparing rows in the table), we see that the rate of 
decline of the mean sSFR$_{\rm gl}$ is much higher for low mass satellites than high mass satellites. 
This further supports the fact that low mass satellites are more sensitive to environment. 
We caution here that the 
analysis above is qualitatively correct as far as we can associate satellites with field galaxies of 
the same present-day stellar mass as their progenitors.
\par
In the literature, the star-formation quenching time-scale is often defined as the time needed for a satellite's 
sSFR to fall below a critical value of 10$^{-11}$ yr$^{-1}$ after infalling into its host. Thus, we repeat 
this approach using our linear fits, and calculate the time it takes to reach this critical value, which 
we dub $\Delta t_{\rm q}$ (in Gyr) in Table 5. We can see that, similar to the rate of decline of sSFR$_{\rm gl}$, 
at fixed stellar mass $\Delta t_{\rm q}$ becomes shorter in more massive environments, once again with the 
unique exception of the most massive stellar mass bin. Now considering fixed halo mass, $\Delta t_{\rm q}$ becomes 
instead {\it{shorter}} for more massive satellite. This is the opposite to what we found for the rate of decline 
of the sSFR$_{\rm gl}$ which happens {\it{more slowly}} in massive satellites. This is, once again, a result of 
the fact that the high mass satellites have low sSFR$_{\rm gl}$ values even prior to infall into their host. 
For example, in the table we can see that satellites in the highest stellar mass bin have a typical value of 
mean sSFR$_{\rm gl}$=1.8$\times$10$^{-11}$ yr$^{-1}$. Given that 1.0$\times$10$^{-11}$ yr$^{-1}$ is the critical 
value for considering galaxies as quenched, this means that they are very close to being quenched even prior to 
infall into their host. So, despite their slow decrease in sSFR$_{\rm gl}$, they require only a short time to 
reach the critical value to be considered quenched. Hence, although at first sight this appears to be a contradiction, 
massive galaxies do actually have the slowest declining sSFR$_{\rm gl}$ inside their present-day host, while 
simultaneously having the shortest values of quenching time-scale.
\par 
It is interesting to compare the $\Delta t_{\rm q}$ values in Table 5 with those derived in the literature using, 
for example, semi-analytic models of galaxy formation and evolution (SAMs). 
De Lucia et al. (2012) compared the observed fractions of passive galaxies at $z \sim$ 0 and their dependence on 
M$_{\rm h}$ and cluster-centric distance with SAMs predictions. They found that satellites with log(M$_{\star}$) $<$
10.5 terminate their star formation activity after having spent 5 - 7 Gyr in haloes more massive than 
log(M$_{\rm h}$) = 13, while more massive satellites require about 5 Gyr. With a similar approach, based on
observed fractions of quiescent satellites, Hirschmann et al. (2014) estimated a quenching time-scale of
$\sim$6 Gyr for low mass galaxies decreasing to $\sim$3 Gyr for satellites more massive than 10$^{10}$ M$_{\odot}$. 
These results are also consistent with the SAM predictions of De Lucia, Hirschmann \& Fontanot (2018).
Wetzel et al. (2013) derived a total quenching time-scale of $\sim$4.5 Gyr to $\sim$6 Gyr at log(M$_{\star}$) $<$ 10
and for decreasing M$_{\rm h}$, between $\sim$3 Gyr and $\sim$5 Gyr in the range 10 $<$ log(M$_{\star}$) $<$ 10.5, 
and finally $\sim$2 - 3 Gyr at higher M$_{\star}$. Using $N-$body cosmological simulations, Oman $\&$ Hudson (2016) 
required a quenching time-scale of $\sim$5 Gyr on average to reproduce the observed fractions of passive satellites, 
nearly independent of M$_{\rm h}$. Thus our $\Delta t_{\rm q}$ values, derived from the observed dependence of the mean 
sSFR$_{\rm gl}$ on projected phase-space, are in good agreement with previous studies, both in terms of their value and 
their dependence on stellar and halo mass. We emphasise our finding that short quenching time-scales (as defined in the 
typical way in the literature) are not necessarily correlated with a rapid decrease in star formation 
rate once a satellite enters the host halo.

\subsubsection{Pre-processing}
As discussed by Berrier et al. (2009), McGee et al. (2009), De Lucia et al. (2012), Lisker et al. (2013),
Vijayaraghavan \& Ricker (2013) and Wetzel et al. (2013) among the  others, a significant fraction of simulated galaxies 
(> 40$\%$ at log(M$_{\star}) \lesssim$ 10.5) were accreted onto present-day cluster mass hosts as satellites of group 
mass haloes. Such galaxies are referred to as pre-processed, in the sense that they may have had their properties transformed 
to some degree prior to infall into their present-day environment. The occurrence of pre-processing in the simulations could 
thus provide a natural explanation for the high fraction of passive satellites observed in massive environments (e.g. van 
den Bosch et al. 2008, Wetzel et al. 2012) and at cluster-centric distances as large as 3R$_{200}$ (Hou et al. 2014, Just et 
al. 2015) without advocating a higher quenching efficiency in cluster mass hosts (cf. also 
Wetzel et al. 2013, Vijayaraghavan \& Ricker 2013). Further observational evidence comes from  
sub-structures identified in the velocity field of galaxy clusters, 
such as Virgo, Fornax, Coma, Abell 85, Abell 2744 and Hydra A/A780, which have been interpreted as accreted lower-mass 
groups (e.g. Fitchett \& Webster 1987, Colless \& Dunn 1996, Petrosian et al. 1998, Zabludoff \& Mulchaey 1998, Drinkwater et 
al. 2001, Edwards et al. 2002, Adami et al. 2005, Boselli et al. 2014, De Grandi et al. 2016, Jauzac et al. 2016, 
Oldham \& Evans 2016, Yu et al. 2016, Lisker et al. 2018). 
\par
There are two features in the trends of Fig. 11 suggestive of pre-processing. Recent infallers, associated with higher-number
zones, have lower global sSFR and older ages than field galaxies of the same stellar mass. This means that galaxies accreted 
a couple of Gyr ago or less are quenched with respect to the field population (despite the high fraction of interlopers). 
In principle, 2 - 3 Gyr are sufficient for a satellite to perform a pericentric passage; our simulations indicate that 
the fraction of recent infallers having experienced a first pericentric passage, which may have partially quenched them,  
is $\sim$40$\%$. This fraction is an upper limit, and would be further reduced if the effects of interlopers were included.
Thus such objects are expected to contribute less to the average galaxy properties in zones 7 and 8 than satellites which 
have yet to reach pericentre (see f$_{\rm no-per}$ in Table 2). It is unlikely that the latter 
have already experienced significant quenching induced by their present-day host halo, and thus the older age and lower 
sSFR$_{\rm gl}$ measured in zone $>$ 6 are likely the result of pre-processing occurred in previous environment(s). 
Additionally, we note that the age difference with the field galaxies is highest for low mass satellites, in particular for
those residing in the most massive haloes. This would be consistent with the results of McGee et al. (2009) and De Lucia et al. 
(2012), which show that massive haloes grow through accretion of substructures comparable to group-mass environments 
where satellites are being quenched. 
\par\noindent
A possible additional evidence for pre-processing is especially visible in the middle panels of Fig. 11, where the distribution
in sSFR$_{\rm gl}$ of satellites in haloes more massive than log(M$_{\rm h}$ M$_{\odot}$ h$^{-1}$) = 13 flattens at zone $>$ 6,
still being lower than the field's sSFR$_{\rm gl}$. At the same time, the distribution in Age$_{\rm L}$ of the same satellites 
levels off at zone $>$ 6 settling around an age older than that of the field. We interpret this feature as due to the recent 
accretion of satellites whose star formation activity already experienced quenching in their previous host environment. 
With regard to this, we checked whether the values at which the distributions in sSFR$_{\rm gl}$ and Age$_{\rm L}$ flatten
match the satellite population from the next M$_{\rm h}$ bin down. 
This is visible in some, but not all, cases, and is difficult to assess given the size of the error bars in 
zone $>$ 6 for the sSFR$_{\rm gl}$ and Age$_{\rm L}$.

\section{Summary}
We combined the group catalogue by Wang et al. (2014), the catalogue of stellar ages and metallicites 
by Gallazzi et al. (2018, in prep)
and the catalogue of global, specific star formation rates by Brinchmann et al. (2004), all drawn for the SDSS DR7 galaxies.
We used these to study the physical properties of satellite galaxies in the projected phase-space of their host environment, 
for satellites found within one virial radius of their host. We limit our analysis to 1R$_{\rm 200}$ because simulations 
show the true cluster population to always dominate over interlopers, and in this way we hope to minimize the effects of 
interlopers on our results.

Using the YZiCS hydrodynamical simulations, we defined a set of zones in projected phase-space, where each zone 
follows the shape of contours of mean infall time ($\overline{\rm T}$$_{\rm inf}$), i.e. the time since a satellite 
galaxy first crossed the virial radius of the main progenitor of its present-day, host halo. 
As the zone number increases from the inner to the outer regions of the projected phase-space diagram, 
we find less ancient infallers and increasing numbers of recent infallers.

We catalogued the zone where each observed satellite is found in projected phase-space (i.e. based on their normalised radial 
velocity ($\Delta$V/$\sigma$) and normalised projected distance (R$_{\rm proj}$/R$_{200}$), measured with respect to the host).
We then analyzed how various properties of the satellites depend on zone and $\overline{\rm T}$$_{\rm inf}$ including: the global 
specific star formation rate (sSFR$_{\rm gl}$), the stellar age (Age$_{\rm L}$), metallicity (log(Z/Z$_{\odot}$)) and 
[$\alpha$/Fe] ratio. Our sample includes satellites with stellar masses from dwarfs up to giants, and hosts with halo masses 
from small groups up to 
massive clusters. As such, we are able to bin our sample by stellar mass (M$_{\star}$) and host halo
(M$_{\rm h}$) mass, in order to cleanly separate environmental quenching from mass quenching. In this way, using our newly 
defined zones, we can study how systematically shifting the mean of the infall time distribution impacts on galaxy properties, 
while simultaneously controlling for the effects of both galaxy mass and host halo mass.
The use of zones in projected phase-space makes it possible to highlight the dependence of galaxy properties on 
environment.
\par\noindent
Our main results can be summarized as follows:
\begin{itemize}
\item {Time since infall is as crucial a parameter as galaxy stellar mass and galaxy environment for understanding
galaxy evolution.} 
\item {At fixed stellar mass and host's halo mass, the luminosity-weighted stellar age and the global, specific
star-formation rate of satellites are sensitive to the time spent in their host environment. This especially holds
for low mass satellites in massive hosts.}
\item {The satellites' average stellar metallicity and [$\alpha$/Fe] abundance ratio appear to depend mostly
on satellite's stellar mass than on host's halo mass. Nevertheless, we see these properties to shift to higher 
values for the ancient infaller population.} 
\item {By considering the ancient infallers alone, we can clearly see the impact of much lower mass hosts, 
meaning that the use of phase-space enables us to more sensitively detect environmental effects.}
\item {When inspecting the dependence of the global, specific star-formation rate on infall time, we see
that the mean sSFR$_{\rm gl}$ declines more rapidly for low mass satellites than high mass satellites. This
is further evidence for low mass satellites being more sensitive to environment on one hand, and for massive
satellites being already or partially quenched upon infall on the other.}
\item {We use the dependence of mean sSFR$_{\rm gl}$ on $\overline{\rm T}$$_{\rm inf}$ to estimate the
quenching time-scale.  We see it varying between $\sim$6 Gyr and $\sim$4 Gyr with increasing host mass for low and 
intermediate mass satellites, while it settles to a value of 2 - 3 Gyr independent of M$_{\rm h}$ for the more massive 
galaxies.} 
\item {Finally, we note that the offset in sSFR$_{\rm gl}$, Age$_{\rm L}$ and log(Z/Z$_{\odot}$) between recent infallers 
and field galaxies, as well as the flattening of the Age$_{\rm L}$ distribution at more recent $\overline{\rm T}$$_{\rm inf}$, 
could be indicative of pre-processing.}
\end{itemize}
{We caution that the above trends likely underestimate the true dependence of satellite observed properties
on T$_{\rm inf}$, as projection effects in 2D phase-space can enhance the spread in infall time in 
a specific zone, and thus modify the true dependence of satellite properties on T$_{\rm inf}$. We expect this
effect to be constant within a zone, thus allowing us to safely compare different galaxy masses 
and different host masses at fixed zone.
\par
{We interpret the dependence of sSFR$_{\rm gl}$, Age$_{\rm L}$ on zone and $\overline{\rm T}$$_{\rm inf}$ as 
the outcome of strangulation and ram pressure stripping which get more efficient in more massive haloes because of their
deeper potential well and 
denser intracluster medium. We also suggest that the progressively lower sSFR$_{\rm gl}$ and older stellar age  
with increasing $\overline{\rm T}$$_{\rm inf}$, could, in part, arise from the accumulative buildup 
of environmental impact with time and, in part, arise from changes in satellite orbits with time (i.e. decreasing pericentric 
distances and orbital eccentricity). 
The latter would make ancient infallers spend more time in the central region of their host halo,
where environmental effects are expected to be more efficient in quenching star formation.
\par\noindent
With regard to metallicity, we interpret the weak trend of increasing log(Z/Z$_{\odot}$) with $\overline{\rm T}$$_{\rm inf}$
as most likely due to ram pressure stripping, which, by removing the galaxy outskirts, inhibits
radial inflows of metal-poor gas and allows the galaxy central regions to form new stars from metal-rich gas
(cf. Bah\'e et al. 2017). The higher $\alpha$-abundance of the ancient infallers could suggest a more truncated
star formation history. 
\par
As a future follow-up to this work, it will certainly be interesting to verify the definition of zones and their relation
with the mean infall time using a larger sample of galaxy groups and clusters from hydro-dynamic simulations and semi-analytic
models. It will also be interesting to compare their predicted galaxy properties with the correlations between observed galaxy
properties and projected phase-space found in this work, in order to better
constrain the efficiency of quenching (e.g. strangulation, ram pressure and tidal stripping) as a function of galaxy infall
time, galaxy orbital parameters and their time evolution as driven by the mass growth of the host halo.

\section*{Acknowledgements}
A.P. and G.D.L. acknowledge support by Sonderforschungsbereich SFB 881 ``The Milky Way System'' (subproject B5 and
visitor programme) of the German Research Foundation. A.G. and S.Z. acknowledge support from Istituto Nazionale di
Astrofisica (PRIN-SKA 2017 program 1.05.01.88.04).
S.K.Y. acknowledges support from the Korean National Research Foundation (NRF-2017R1A2A05001116). This study was performed 
under the umbrella of the joint collaboration between Yonsei University Observatory and the Korean Astronomy and Space 
Science Institute. The supercomputing time for numerical simulation was kindly provided by Korea Institute of Science and 
Technology Information (KISTI KSC-2014-G2-003), and large data transfer was supported by Korea Research Environment Open Network
(KREONET), which is managed and operated by KISTI. We thank the referee and
Quan Guo for valuable comments that helped improve this paper.






\appendix
\section{Caveats on the observed trends in phase-space}
\subsection{Impact of interlopers}
Each of the trends of satellite observed property with zone is expected to be contaminated with interlopers.
In this case, we classify interlopers as field galaxies that are beyond the virial radius  of the group/cluster, but are simply
projected down the line-of-sight to appear within the zones of the phase-space diagram. We have used the YZiCS simulations to
quantify the fraction of interlopers in each zone: f$_{\rm intl}$ is listed in Table 2 and increases from ~10$\%$ to ~40$\%$
between zone 1 and 8. This means that the true members of a host always dominate over the interlopers in our
phase-space diagrams, and for small zone numbers the interlopers contribute only weakly (less than a 20$\%$ contribution for
zone $<$ 4). In fact, this provides added motivation to apply phase-space diagrams when studying environmental effects.
For example, in Figs. 7 and 8, not only do galaxies in zone $<$ 2 show stronger effects of environment because they have
been subjected to that environment for longer on average, but also because they are much less contaminated by interlopers and
thus the host mass better reflects their true environment.
\par\noindent
However, since field galaxies are typically younger and have higher global sSFR than satellites (cf. Fig. 11), we expect
interlopers to alter the true age/sSFR gradients of group/cluster galaxies in phase-space (this applies also to the
metallicity gradients in Fig. 12). Nevertheless, within a single zone, we can expect a fixed interloper fraction, and thus
comparison between galaxy properties at fixed zone is not affected by this issue. We also note that, although the varying
interloper fraction might alter the gradient with respect to zone, there is little reason to expect this to have a
significant effect on the observed dependencies on host mass. An example of this comes from the changes in Age$_{\rm L}$
along the host mass axis (y-axis) in Figure 8. The changes in this direction that are seen when switching from the zone $<$ 2
panel to the zone $\geq$ 5 panel are unlikely to be strongly effected by interlopers, simply because many interlopers are
field galaxies  actually very distant from the group/cluster. Their properties are not expected to be a strong function of
the mass of a group/cluster that they are projected onto by chance.

\begin{figure*}
\includegraphics[width=150mm]{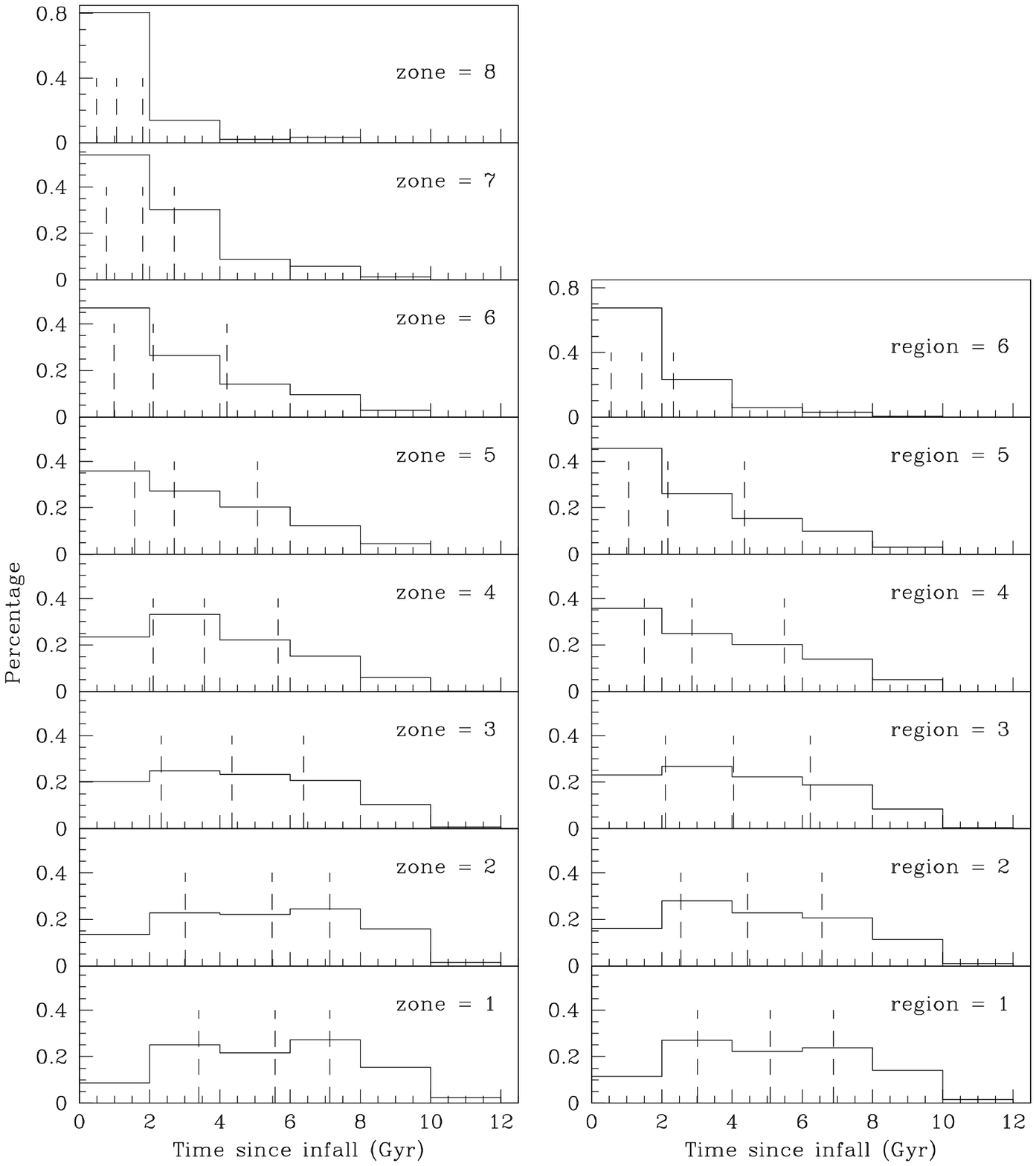}
\caption{The normalized T$_{\rm inf}$ distributions of simulated satellites residing in environments with
13.7 $\leq$ log(M$_{\rm h}$ M$_{\odot}) \leq$ 15, and distinguished among zones (left-hand panels) and caustic
regions (right-hand panels). The dashed lines indicate, from left to right, the 25$\%$, 50$\%$ and 75$\%$
percentiles of the T$_{\rm inf}$ distribution in each zone and caustic region.}
\end{figure*}

\subsection{Spread in infall-time within zones}
Even in a 3D phase-space diagram, that is free from projection effects, there is some spread in infall times at a particular
location in phase-space (see for example the left panel of Fig. 2 from Rhee et al. 2017). This may partly arise due to the
variety of orbital parameters that members of a host may have (see Section 5.1.1 for an analysis of orbital parameters in
phase-space). However, in a 2D phase-space diagram, projection effects significantly increase the spread in infall
time at all locations in phase-space.
\par\noindent
To highlight this, we have used the simulations and plotted in Fig. A1 the normalized
distribution in T$_{\rm inf}$ of satellites belonging to each of the 8 zones and 6 caustic regions
defined in Sect. 3.1. Here, we take into account all simulated satellites and do not apply any cut  in stellar mass.
We also use dashed lines to show the 25$\%$, 50$\%$ and 75$\%$ percentiles of the T$_{\rm inf}$ distribution
in each zone and caustic region. As we move from zone = 1 to zone = 8 in the left-hand plots of Fig. A1, we see
that the peak and percentiles of the distribution shifts to lower T$_{\rm inf}$ and the tail towards larger
T$_{\rm inf}$ drops but does not disappear altogether. For example, in zone 8, which is characterized by
$\overline{\rm T}$$_{\rm inf} =$ 1.42 Gyr and $\sigma$($\overline{\rm T}$$_{\rm inf})$ = 1.49 Gyr, the contaminant
galaxies with T$_{\rm inf} > $ 4 Gyr amount to $\sim$6$\%$ of all satellites. In the opposite direction, from zone 8 to zone 1,
we see the peak of the distributions moving to higher T$_{\rm inf}$, with the tail towards low T$_{\rm inf}$ progressively
decreasing. For example, in zone 1 with $\overline{\rm T}$$_{\rm inf} =$ 5.42 Gyr and $\sigma$($\overline{\rm T}$$_{\rm inf})$
= 2.51 Gyr, the contaminant satellites with T$_{\rm inf} <$ 4 Gyr represent $\sim$34$\%$ of the galaxy population.
\par\noindent
It is interesting to compare the above distributions with those obtained using, instead, regions defined via the caustic profiles
(as drawn in Fig. 4). The latter are plotted in the right-hand column of Fig. A1. Overall, the peak and
percentiles of the distribution shifts from ancient infallers towards recent infallers with increasing curve number, in a
qualitatively similar way as it does with increasing zone number. However, simply from visually inspecting Fig. 4, it is clear that
we might expect slightly more mixing of infall times with the caustic curves than with the zones.
In the case of caustic regions, the fraction of contaminants (satellites with T$_{\rm inf} >$
4 Gyr) for region 6 is $\sim$9$\%$, while the fraction of contaminants (satellites with T$_{\rm inf} <$ 4 Gyr)
in region 1 is $\sim$39$\%$, slightly higher than what derived for zones 8 and 1. We thus gather that our newly-defined
set of zones in projected phase-space not only separates regions of different T$_{\rm inf}$ in a cleaner way than caustic
profiles, but also slightly reduces the spread in T$_{\rm inf}$ within a zone (see also Table 1). None the less, the effect
is not as pronounced as it might appear when comparing the
two panels in Fig. 4. For example, caustic region 1 clearly cuts across regions with quite widely differing infall time.
The reason this does not cause a more significant change in the distribution of infall times is because the  number density of
galaxies is non-uniform across the plot, and quite low where the most differing infall times are located.


\bsp	
\label{lastpage}
\end{document}